\documentclass[12pt]{article}
\pdfoutput=1
\usepackage{amsmath,amssymb,amsthm}
\usepackage{bm}
\usepackage[usenames,dvipsnames]{xcolor}
\usepackage{graphicx}
\usepackage{float}
\usepackage{array}
\usepackage{arydshln}
\usepackage{multirow}
\usepackage{rotfloat}
\usepackage{caption}
\usepackage{subcaption}
\usepackage[normalem]{ulem}
\usepackage{cite}
\usepackage{hyperref}
\usepackage{cleveref}

\numberwithin{equation}{section}
\numberwithin{figure}{section}

\graphicspath{{figures/}}

\theoremstyle{definition}

\begin{document}

\begin{titlepage}
\vspace*{-3cm} 
\begin{flushright}
{\tt CERN-TH-2023-067\\ZMP-HH/23-6}
\end{flushright}
\begin{center}
\vspace{2cm}

{\LARGE\bfseries On Higher-Spin Points and Infinite Distances\\[7pt] in Conformal Manifolds}
\vspace{1.2cm}

{\large
Florent Baume$^{1,2}$ and Jos\'e Calder\'on-Infante$^{3}$\\} 
\vspace{.7cm}
{ $^1$ Department of Physics and Astronomy, University of Pennsylvania}\par
{Philadelphia, PA 19104, U.S.A.}\par
\vspace{.2cm}
{ $^2$ II. Institut f\"ur Theoretische Physik, Universit\"at Hamburg, Luruper Chaussee 149,}\par
{22607 Hamburg, Germany}\par
\vspace{.2cm}
{ $^3$ Theoretical Physics Department, CERN,}\par{ CH-1211 Geneva 23, Switzerland}\par

\vspace{.3cm}

\scalebox{1.0}{\tt florent.baume@desy.de, jose.calderon-infante@cern.ch}\par
\vspace{1.2cm}
\textbf{Abstract}
\end{center}

Distances in the conformal manifold, the space of CFTs related by marginal
deformations, can be measured in terms of the Zamolodchikov metric. Part of the
CFT Distance Conjecture posits that points in this manifold where part of the
spectrum becomes free, called higher-spin points, can only be at infinite
distance from the interior. There, an infinite tower of operators become
conserved currents, and the conformal symmetry is enhanced to a higher-spin
algebra. This proposal was initially motivated by the Swampland Distance
Conjecture, one of pillars of the Swampland Program. In this work, we show
that the conjecture can be tackled using only methods from the conformal
toolkit, and without relying on the existence of a weakly-coupled gravity dual.
Via conformal perturbation theory combined with properties of correlators and
of the higher-spin algebra, we establish that higher-spin points are indeed at
infinite distance in the conformal manifold. We make no assumptions besides the
usual properties of local CFTs, such as unitarity and the existence of an 
energy-momentum tensor. In particular, we do not rely on a specific dimension
of spacetime (although we assume $d>2$), nor do we require the presence of
supersymmetry.

\vfill 
\end{titlepage}

\setcounter{tocdepth}{2}

\tableofcontents

\enlargethispage{\baselineskip}

\newpage

\section{Introduction}

Exactly marginal operators play a special r\^ole in local Conformal Field
Theory (CFT): one can use them to deform the original theory without breaking
spacetime symmetries and reach infinite families of CFTs related by the tuning
of associated coupling constants. These span the so-called conformal manifold,
and are not uncommon in four dimensions and lower. Supersymmetry is however
usually required in order to protect marginal operators against unwanted
quantum corrections and maintain a vanishing beta-function, see e.g. references
\cite{Gates:1983nr, Seiberg:1988ur, Leigh:1995ep, Kol:2002zt} for early works.

While there are examples of non-supersymmetric conformal manifolds in two
dimensions related to the compact free boson and its toroidal orbifold
generalizations, to the best of our knowledge all well-established examples in
higher dimensions are supersymmetric. Furthermore, the possible types of
supersymmetry-preserving exactly marginal deformations being severely
restricted by representation theory, supersymmetric conformal manifolds cannot
exist in dimensions greater than four, see reference \cite{Cordova:2016xhm} for
an exhaustive analysis. Recently, there has been a growing effort to study
conformal manifolds via various techniques in supersymmetric theories
\cite{Kol:2010ub, Chang:2010sg, Green:2010da, Gomis:2015yaa, Buican:2014sfa,
Baggio:2017mas, Beratto:2020qyk, Niarchos:2021iax}, as well as with methods not
relying on such additional structures \cite{Osborn:2015rna, Bashmakov:2016uqk,
Bashmakov:2017rko, Sen:2017gfr,Hollands:2017chb, Behan:2017mwi,
Balthazar:2022hzb}.

Conformal manifolds can allow for special points where a sector of the CFT
decouples from the rest of the spectrum and becomes free. At these points, an
infinite tower of higher-spin operators become conserved currents, and the
conformal symmetry enlarges to an infinite-dimensional higher-spin
algebra.\footnote{By higher-spin operators, we will always mean operators
transforming in the $\ell$-traceless-symmetric representation of the Lorentz
group.} These symmetry-enhanced points, dubbed higher-spin (HS) points, can
tell us much about the underlying physics, as it was shown that the converse is
also true: the presence of a higher-spin symmetry always implies that at least
part of the theory is free. First shown in $d=3$ by Maldacena and Zhiboedov
\cite{Maldacena:2011jn}, this theorem was swiftly generalized to arbitrary
dimensions \cite{Stanev:2013qra, Boulanger:2013zza, Alba:2013yda, Alba:2015upa,
Hartman:2015lfa, Li:2015itl}. Moreover, a direct corollary is that the presence
of a single higher-spin conserved current is a sufficient condition to obtain
infinite tower of such operators, as it is essentially a consequence of the
closure of the commutation relations of the algebra generators. Hunting for an
operator becoming a conserved current in a certain limit can therefore help us
find dual descriptions in terms of a weakly-coupled sector.

A conformal manifold comes furthermore equipped with the Zamolodchikov metric,
defined from the two-point function of the marginal operators. The properties
of correlators in unitary theories then ensure that it has the expected
properties of a metric, and enables one to define a notion of distance between
different CFTs related by marginal deformations.

Given both the potential presence of symmetry-enhanced points and the ability
to measure distances between theories, it is worthwhile to find universal
features of conformal manifolds, perhaps relating them. For instance, one can
ask about those CFTs that lie at infinite distance from generic points. It is
then natural to expect that they correspond to physically-distinguished
theories. Such special cases occur, for example, at points where there is a
symmetry enhancement. In particular, one may wonder whether there is a
connection between infinite distances and higher-spin enhancements in the
conformal manifold.

\medskip

Through the AdS/CFT correspondence, questions about universal properties of
CFTs are very much in the spirit of the Swampland Program \cite{Vafa:2005ui},
see e.g. references \cite{Brennan:2017rbf, Palti:2019pca, vanBeest:2021lhn,
Grana:2021zvf} for reviews. In the bulk, the conformal manifold maps to the
space of vacuum expectation values of massless scalar fields, i.e. the moduli
space of the gravity theory. Within the scope of the Swampland Program,
considerable effort has been recently spent with the goal of probing and
studying the structure of these spaces, and a number of conjectures pointing to
universal behaviors have been formulated. On the field-theory side, this can serve as a
guide to find universal properties of conformal manifolds.

For instance, with the advent of the AdS/CFT correspondence the importance of
HS points was promptly recognized, for instance in the duality between vector
models and gravitational theories of higher-spin fields \cite{Klebanov:2002ja}
and its generalizations, see e.g. reference \cite{Giombi:2016ejx} for a review.
Moreover, the breaking of this symmetry in $\mathcal{N}=4$ super-Yang--Mills
can be understood holographically as realizing the Pantagruelic Higgs mechanism
(``La Grande Bouffe'') responsible for giving a mass to higher-spin fields in
the $\text{AdS}_5\times S^5$ bulk theory \cite{Bianchi:2003wx, Beisert:2003te,
Beisert:2004di}.

In the same vein, there is a prominent Swampland conjecture pointing to
universal behaviors of infinite-distance points in the moduli space: the
Swampland Distance Conjecture \cite{Ooguri:2006in}. It states that as one
approaches a point that is at infinite distance from the interior, an infinite
tower of states become massless, and does so exponentially fast with the
distance, i.e. $M_\text{tower}\sim e^{-\alpha_G\cdot \text{dist}}$ in Planck
units. These points are further expected to occur as special limits, and are
related either to a decompactification, where the tower corresponds
Kaluza--Klein modes, or a limit for which a fundamental string becomes weakly
coupled, a proposal called the Emergent String Conjecture \cite{Lee:2019xtm}.

The Swampland Distance Conjecture enjoys a large body of evidence in string
theory compactifications leading to flat space backgrounds, in particular in
supersymmetric setups \cite{Ooguri:2006in, Blumenhagen:2017cxt, Grimm:2018ohb,
Lee:2018urn, Lee:2018spm, Grimm:2018cpv, Buratti:2018xjt, Corvilain:2018lgw,
Lee:2019tst, Joshi:2019nzi, Marchesano:2019ifh, Font:2019cxq, Lee:2019xtm,
Lee:2019wij, Baume:2019sry, Cecotti:2020rjq, Gendler:2020dfp, Lee:2020gvu,
Lanza:2020qmt, Klaewer:2020lfg, Lanza:2021udy, Palti:2021ubp,
Etheredge:2022opl}. Despite the amount of support in favor of the conjecture,
one could worry that it is merely coming from two lamppost effects: string
theory and supersymmetry. It is however challenging to go beyond these regimes
due to the lack of a general framework describing any theory of quantum
gravity, and the lack of technical tools in non-supersymmetric setups.

A promising way out is the holographic approach. This framework is \emph{a
priori} independent of string theory, and there are powerful, non-perturbative,
approaches in the conformal toolkit to go beyond supersymmetry. In this way,
not only can the Swampland Program serve as guidance for questions related to
conformal field theories, but methods such as the conformal bootstrap can help
in the gathering of evidence for Swampland conjectures going beyond lamppost
effects. This ``holographic Swampland Program'' has in fact already bore fruits
when applied to the absence of global symmetries in quantum gravity \cite{Harlow:2018tng,
Harlow:2018jwu} the Weak Gravity Conjecture \cite{Montero:2016tif,
Heidenreich:2016aqi, Urbano:2018kax, Montero:2018fns, Aharony:2021mpc,
Palti:2022unw, Andriolo:2022hax, Sharon:2023drx, Orlando:2023ljh}, moduli
stabilization, and the scale-separation problem \cite{Conlon:2020wmc,Conlon:2021cjk,
Apers:2022tfm, Plauschinn:2022ztd, Quirant:2022fpn, Apers:2022vfp,
Montero:2022ghl}.

\medskip

For the Distance Conjecture this approach originated in references
\cite{Baume:2020dqd,Perlmutter:2020buo}. There, it was found that in large
classes of supersymmetric CFTs (SCFTs) the Swampland Distance Conjecture is
naturally realized in the conformal manifold as one approaches a higher-spin
point. The prototypical example is that of four-dimensional $\mathcal{N}=2$
SCFTs, for which marginal couplings correspond to complexified gauge
parameters. Around the weak-coupling point, $\tau \to i \infty$, the
Zamolodchikov metric turns out to be approximately hyperbolic,
$\chi_{\tau\bar{\tau}}\sim(\text{Im}\tau)^{-2}$ \cite{Gerchkovitz:2016gxx}. The
geodesic distance growing logarithmically with $\tau$, the free point is at
infinite distance. Combined with the fact that HS currents have an anomalous
dimension given at one loop by $\gamma=\Delta-\Delta_\text{free} \sim
(\text{Im} \tau)^{-1}$, they become conserved exponentially fast with the
Zamolodchikov distance. 

This process is natural in supersymmetric gauge theories and leads to the
intriguing possibility that it is in fact the general mechanism behind the
Swampland Distance Conjecture from an holographic perspective. In reference
\cite{Perlmutter:2020buo}, this expectation was encapsulated in the CFT
Distance Conjecture.

\paragraph{CFT Distance Conjecture:} given the conformal manifold of a local CFT in $d>2$,
\begin{itemize}
	\item[] \textbf{Conjecture I:} all higher-spin (HS) points are at infinite distance;
    
	\item[] \textbf{Conjecture II:} all CFTs at infinite distance are HS points;
    
	\item[] \textbf{Conjecture III:} the anomalous dimensions of the
			higher-spin operators becoming conserved currents at the HS point go to zero exponentially fast with the geodesic distance:
			\begin{equation} \label{exponential-SDC}
					\gamma = \Delta - \Delta_\text{free}  \sim e^{-\alpha_\chi \text{dist}_\chi}\,.
			\end{equation}
\end{itemize}
The distances are computed with respect to the Zamolodchikov metric, and the
coefficient $\alpha_\chi$ is expected to be related to the central charge of
the theory, $C_T$. In particular, $\sqrt{C_T} \, \alpha_\chi$ should at least
of order one \cite{Baume:2020dqd,Perlmutter:2020buo}. By local CFT, we further
mean theories whose spectrum include the energy-momentum tensor. This condition
is very relevant from the Swampland perspective, since it translates to
dynamical gravity in the bulk. Moreover, this conjecture relies on the
conformal manifold being parameterized by exactly marginal operators, which is
required to have a well-defined notion of distance in terms of the
Zamolodchikov metric. For instance, there are theories without local
energy-momentum tensor that preserve conformal invariance for any value of a
continuous parameter, but that do not have an associated local marginal
operator, see e.g. references \cite{Behan:2017emf, DiPietro:2019hqe}.

Furthermore, note that the conjecture does not claim that any CFT should have a
conformal manifold. For instance, in five and six dimensions, there are no
supersymmetry-preserving marginal deformations, and it is believed that all 5d
and 6d CFTs are supersymmetric isolated points. In addition, it is not required
that all conformal manifolds must contain infinite-distance points. There are
indeed known cases of compact conformal manifolds \cite{Buican:2014sfa}, and
superpotential deformations in $\mathcal{N}=1$ four-dimensional SCFTs are
expected to always lead to finite distances \cite{Leigh:1995ep, Green:2010da,
Perlmutter:2020buo}. Due to the close resemblance between 4d $\mathcal{N}=1$
and 3d $\mathcal{N}=2$ SCFTs, it is further predicted that conformal manifold
of three-dimensional theories should always be compact
\cite{Perlmutter:2020buo}.

In two dimensions the status of the conjecture is unclear, as both the global
conformal groups and the Virasoro algebra always admits higher-spin currents.
Moreover, there also exists HS algebras that are finitely generated. A possible
extension of the CFT Distance Conjecture in two dimensions could come from the
fact that limiting points where the scalar gap vanishes, $\Delta\to0$, are at
infinite distance \cite{Perlmutter:2020buo}. Indeed, it was similarly
conjectured that the converse is also true \cite{Kontsevich:2000yf,
Acharya:2006zw}. Due to these complications, we will however restrict ourselves
to cases where the spacetime dimension is $d>2$.

\medskip

As discussed above, the evidence for the conjecture mainly comes from SCFTs.
Despite this, as stated the CFT Distance Conjecture does not make any
assumption on the existence of supersymmetry nor does it differentiate between
dimensions. This suggests that the mechanism behind this feature of conformal
manifolds should not rely on supersymmetry, regardless of whether there are
conformal manifold associated with non-supersymmetric CFTs. If this is to be
the case, there should exist a way to prove at least part of the conjecture in
a way that is independent of the dimension and without assuming additional
ingredients. This is quite reminiscent of many of the tools used in the study
of CFTs. A celebrated example is the set of equations constraining the form of
the conformal block expansion of four-point functions---the keystone of the
modern conformal bootstrap---which vary continuously with the dimension
\cite{Dolan:2000ut,Dolan:2003hv}. Supersymmetry then merely imposes additional
selection rules between the blocks. In the context of the holographic Swampland
program, a general result in this direction can be found in references
\cite{Stout:2021ubb,Stout:2022phm}, where infinite distance limits have been
linked to the factorization of correlators in the theory. For instance,
this happens at HS points that contain a scalar operator saturating the
unitarity bound, $\Delta=(d-2)/2$, and thus corresponding to a free scalar
field. 

In this work, we will show that it is indeed possible to prove for Conjecture I
in such a way, i.e., only assuming that the CFT is local and possesses a
conformal manifold. The crux of our argument relies on the constraints imposed
by HS symmetry on correlators involving higher-spin currents.

We will use that, away from a given reference CFT, the changes in the physical
data can be tracked in terms of conformal perturbation theory
\cite{Zamolodchikov:1987ti}. For instance, scaling dimensions can be obtained
from three-point functions involving a relevant operator deforming the theory.
Such techniques have been for instance utilized to explore the space of
theories beyond the conformal regime and (re)discover new fixed points, see
e.g. references \cite{Osborn:2017ucf, Rychkov:2018vya, Codello:2020lta,
Osborn:2020cnf, Hogervorst:2020gtc, Antunes:2022vtb} for recent advances in that direction. 

Here, we will focus on exactly marginal deformations to travel in the conformal
manifold, where changes in the conformal dimensions away from a reference
point are encoded in three-point functions involving marginal operators.
Deforming away from the HS point we can then compute anomalous dimensions
at leading order and relate them to the Zamolodchikov distance in a way that does
not require a Lagrangian description, or a precise microscopic description of
the marginal operators, only their existence.

\subsection{Sketch of the Proof}

As our proof will invoke conformal correlators involving higher-spin operators,
as well as sometimes technical properties of higher-spin algebras and the
induced conservation relations, we now summarize the steps needed to prove the
first part of the CFT Distance conjecture, namely that all higher-spin (HS)
points are at infinite distance in the conformal manifold.

The relevant data, the conformal dimension $\Delta_\ell$ of spin-$\ell$
operators $J_\ell$, is encoded in their two-point functions. Choosing a
trajectory corresponding to a coordinate $t$ in the conformal manifold
$\mathcal{M}$, one can use conformal perturbation theory to show that
$\Delta_\ell(t)$ is controlled by the following differential equation
\cite{Sen:2017gfr}:
\begin{equation}
		\frac{d\Delta_\ell}{dt} = - C_{JJ\mathcal{O}}(t)\,,
\end{equation}
where $C_{JJ\mathcal{O}}$ is a specific combination of the coefficients
appearing in the correlator $\left<J_\ell J_\ell\mathcal{O}\right>$. Since the
coefficients of the three-point functions are related to those appearing in
Operator Product Expansions (OPEs), we will often use a shorthand, and refer to
$C_{JJ\mathcal{O}}$ as the ``OPE combination''.

The behavior of the anomalous dimensions near a reference point of
$\mathcal{M}$ therefore depends on how $C_{JJ\mathcal{O}}$ varies in that
neighborhood. In particular, if for any marginal operator $\mathcal{O}$ with
properly normalized two-point function, we have
$C_{JJ\mathcal{O}}\sim\gamma^{1+a}$ with $a\geq0$ close to a HS point, then the
above differential equation implies the anomalous dimension vanishes at
$\text{dist}_\chi\sim t\to\infty$. That is, the HS point is at infinite
distance from any other point.

To show this we introduce an expansion parameter $g$, related to the coordinate
$t$, measuring the breaking of the conservation equation for the HS
currents---that is, the divergence no longer vanishes:
\begin{equation}\label{broken-ward-intro}
		\partial\cdot J_\ell = g\,K_{\ell-1}\,,
\end{equation}
where $K_{\ell-1}$ is a spin-$(\ell-1)$ operator that depend on the microscopic
details of theory, and has conformal dimension $\Delta = d - 1 +\ell$ at the
higher-spin point. This will allow us to compute the OPE combination as a
perturbative series in $g$ of the form
\begin{equation}
		C_{JJ\mathcal{O}} = C_{JJ\mathcal{O}}^{(0)} + C_{JJ\mathcal{O}}^{(1)}\,g + C_{JJ\mathcal{O}}^{(2)}\,g^2 + \cdots \, .
\end{equation}
Most importantly, this can be rewritten in terms of quantities evaluated at the
HS point using \emph{Anselmi's trick} \cite{Anselmi:1998ms}: by taking
different numbers of divergences on $\left<J_\ell J_\ell\mathcal{O}\right>$,
and employing the broken Ward identity given in equation
\eqref{broken-ward-intro}, each coefficient in the expansion above gets
naturally related to three-point functions involving different numbers of the
operator $K_{\ell}$, \emph{evaluated at the higher-spin point}:
\begin{equation}
		C_{JJ\mathcal{O}} = C_{JJ\mathcal{O}}^\text{HS} + C_{JK\mathcal{O}}^\text{HS}\,g + C_{KK\mathcal{O}}^\text{HS}\,g^2 + \cdots\,.
\end{equation}
Here $C_{JK\mathcal{O}}$ and $C_{KK\mathcal{O}}$ denote certain linear
combinations of the OPE coefficients appearing in the correlators $\left<J_\ell
K_{\ell-1} \mathcal{O}\right>$ and $\left<K_{\ell-1}
K_{\ell-1}\mathcal{O}\right>$, respectively. 

Using the constraints imposed by HS symmetry on these correlators, we will show
that $C_{JJ\mathcal{O}}^\text{HS} = 0 = C_{JK\mathcal{O}}^\text{HS}$ for any
marginal operator. The first non-trivial contribution to $C_{JJ\mathcal{O}}$
therefore comes at order $g^2$ or higher. An application of Anselmi's trick on
the two-point functions of $J_\ell$ is furthermore well known to show that at
leading order we have the relation $\gamma \sim g^2$, and we finally obtain
that in the neighborhood of a HS point, we must indeed have:
\begin{equation}
	C_{JJ\mathcal{O}}\sim \gamma^{1+a}\quad \text{ with } a\geq0 \,,
\end{equation}
for any marginal operator. We then conclude that any point of the conformal
manifold with a higher-spin symmetry must be at infinite distance.

The rest of this work fills in the details and is organized as follow: in
Section \ref{sec:finite-vs-infinite-with-CPT}, we review the structure of two-
and three-point correlators involving higher-spin operators as well as
conformal perturbation theory, and give a criterion to decide whether a point
of the conformal manifold is at finite or infinite distance. In Section
\ref{sec:HS-at-infinite-distance} we show that HS points are at infinite
distance following the procedure sketched above, relegating the more technical
computations to the appendices. In Section \ref{sec:flavor-infinite-distance}
we briefly discuss the fate of points where a new flavor conserved current
appear in the spectrum. We give our conclusions in Section
\ref{sec:conclusions}.

\section{Distances from Conformal Perturbation Theory}\label{sec:finite-vs-infinite-with-CPT}

To characterize limiting behavior in the conformal manifold, we need to
understand how the relevant conformal data changes under marginal deformations.
For our purpose this is the conformal dimension $\Delta_\ell$ of higher-spin
operators, as HS points are characterized by a tower of operators saturating
unitarity bounds. As we will review in this section, the changes are encoded in
an evolution equation relating the conformal dimension to the coefficients of a
particular set of three-point functions. This is possible due to the severe
constraints imposed by conformal symmetry, which completely fixes the kinematic
structure of the three-point functions up to these coefficients. 

In this section, we review those constraints, as well as how correlators vary
under conformal perturbation theory. Working in the neighborhood of a reference
point of the conformal manifold, this will then enable us to find a criterion
allowing us to decide whether a point is at
finite or infinite distance.

\subsection{Spinning Conformal Correlators}\label{sec:spinning-correlators}

The central objects of this work are higher-spin (HS) operators $J_\ell$
becoming conserved currents at the HS point. Away from this special point they
develop an anomalous dimension and are no longer conserved, but remain
conformal primaries transforming in the $\ell$-traceless-symmetric
representation of the $d$-dimensional Lorentz group, i.e. spin-$\ell$
operators. When studying such operators, it is convenient to use an index-free
notation by contracting all Lorentz indices with polarization vectors,
$\xi^\mu$:\footnote{For the sake of clarity, we will often drop the dependence
of the operators on their polarization vectors when there are no ambiguities,
leaving it implicit, e.g. $J_\ell(x)=J_\ell(x,\xi)$.}
\begin{equation}
		J_\ell(x,\xi) = J_{\ell,\,\mu_1\dots\mu_\ell}\,\xi^{\mu_1}\cdots \xi^{\mu_\ell}\,.
\end{equation}

The structure of correlators involving higher-spin operators was studied in
detail in references \cite{Giombi:2011rz,Costa:2011mg}, where it was shown that
the tensor structures allowed by conformal invariance can be constructed out of
simple building blocks. For instance, the two-point function of two spinning
operators is constrained to have the following form:
\begin{equation} \label{2pt-function}
		\left<J_\ell(x_1,\xi_1)J_\ell(x_2,\xi_2)\right> = \frac{H_{12}(x_{12};\xi_1,\xi_2)^\ell}{|x_{12}|^{2\tau}} \, , \quad 
\end{equation}
where we defined $x_{ij}=x_i-x_j$, and we use a convention where the
denominator is given in terms of the twist:
\begin{equation}
    \tau = \Delta_\ell - \ell \, .
\end{equation}
In this case, there is only a single kinematic function, $H_{12}$, respecting
conformal covariance. It is related to the well-known spin-one projector
appearing e.g. in scattering amplitudes involving gauge bosons, and its
explicit expression can be found in equation \eqref{building-blocks}. Note that
we will be working in an orthonormal basis for a given spin $\ell$, thereby
fixing the prefactor of the two-point functions to one.

On the other hand, three-point functions involving \emph{a priori} different
spins, which we will refer to as a type-$(\ell_1,\ell_2,\ell_3)$ correlators,
have more than a single kinematic structure allowed by symmetry, and can be
constructed out of six different building blocks \cite{Costa:2011mg}. In this
work, we will only deal with type-$(\ell, \ell, 0)$ and type-$(\ell,\ell-1, 0)$
correlators, leading to a simplification of the allowed structures, as they can
be constructed out of only on three building blocks. These include the
aforementioned kinematic function $H_{12}$, as well two additional blocks,
denoted $V_1\,,V_2$. Their precise form in terms of spacetime coordinates and
polarization vectors will be relevant only in a few technical steps performed
in the appendices. In the main text, we will only need to know that they allow
us to construct a basis of conformally-covariant structures, and we have
therefore relegated their definitions to Appendix \ref{app:surface-integrals}.

\medskip

As we will see shortly, a key r\^ole will be played by the three-point
functions of two higher-spin operators and a marginal operator $\mathcal{O}$.
This is a type-$(\ell,\ell,0)$ correlator, generically taking the following
form:
\begin{equation}\label{conformal-structure-expansion}
		\left<J_\ell(x_1,\xi_1)J_\ell(x_2,\xi_2)\mathcal{O}(x_3)\right> = \sum_{n=0}^\ell C_{JJ\mathcal{O}}^{\,n} \,\Theta_n(x_i;\xi_1,\xi_2). 
\end{equation}
There are $\ell+1$ independent conformal structures $\Theta_n$ that depend both
on Lorentz coordinates and the polarization vectors of the two higher-spin
operators. In terms of the building blocks, these conformally-covariant
functions are given by
\begin{equation}\label{conformal-structure}
		\Theta_n(x_i;\xi_1,\xi_2) = \frac{H_{12}^{\ell -n}\big(V_1 V_2\big)^{n}}{|x_{12}|^{\tau_{123}}|x_{13}|^{\tau_{132}}|x_{23}|^{\tau_{231}}}\,,\qquad \tau_{ijk} = \tau_i + \tau_j - \tau_k\,.
\end{equation}
The numbers $C_{JJ\mathcal{O}}^{\,n}\in \mathbb{R}$ are related to the Operator
Product Expansion (OPE) and therefore depend on the microscopic details of the
theory under consideration, and we often refer to them as the OPE coefficients.

We will also be led to consider three-point functions of a marginal operator
and two different operators $J_\ell$ and $K_{\ell-1}$ of adjacent spins. For
those correlators of type $(\ell,\ell-1,0)$, one finds a very similar
expression to that of equation \eqref{conformal-structure-expansion}:
\begin{equation} \label{JKO}
	\left<J_{\ell}(x_1,\xi_1) K_{\ell-1}(x_2,\xi_2) \mathcal{O}(x_3) \right> = \sum_{n=0}^{\ell-1} C_{JK\mathcal{O}}^{\,n}\, \widetilde{\Theta}_{n}(x_i;\xi_1,\xi_2) \, ,
\end{equation}
where there are now $(\ell-1)$ independent conformal structures given in terms
of the building blocks by:
\begin{equation}\label{conformal-structure-tilde}
		\widetilde{\Theta}_{n}(x_i;\xi_1,\xi_2) = \frac{H_{12}^{\ell -n-1}V_{1}^{n+1} V_2^{n}}{|x_{12}|^{\tau_{123}}|x_{13}|^{\tau_{132}}|x_{23}|^{\tau_{231}}} \,,\qquad \tau_{ijk} = \tau_i + \tau_j - \tau_k\,. 
\end{equation}

Note that as these two cases illustrate, contrary to the usual intuition coming
from three-point functions of scalar operators where everything is fixed up to
a single coefficient, there can be more than one when dealing with spinning
correlators. As we will see below, there can however be relations between them
in a given theory, and this will be one of the ingredients of our proof. For
instance, conservation conditions reduce the number of independent conformal
structures, leading to constraints on the OPE coefficients

\medskip

We close by commenting on the special case where $d=3$. The conformal
structures that we have reviewed are parity even. For $d\geq 4$ these are all
the structures that can appear in these correlators, but in three dimensions
there are $\ell$ and $\ell-1$ extra parity-odd structures allowed to appear in
$\left<J_\ell J_\ell\mathcal{O}\right>$ and $\left<J_\ell
K_{\ell-1}\mathcal{O}\right>$, respectively. Nevertheless, they will not play
any r\^ole in our analysis. First, parity-odd structures will not appear in
conformal perturbation theory applied to conformal dimensions, and second,
there is no mixing between parity-odd and parity-even structures in the
constraints imposed by HS symmetry. We will come back to this point in sections
\ref{sec:CPT} and \ref{sec:charge-conservation}.

\subsection{Conformal Perturbation Theory} \label{sec:CPT}

As we will now review, the usefulness of two- and three-point functions reveals
itself in the study of conformal perturbations \cite{Zamolodchikov:1987ti},
allowing one to track the evolution of the conformal data as the parameters of
the theory are tuned. In particular, since exactly marginal deformations do not
break conformal invariance and span the conformal manifold $\mathcal{M}$, we
can use conformal perturbation theory around a reference point of $\mathcal{M}$
to, at least formally, follow the modifications of the conformal data as a
function of the couplings.

Placing ourselves at a given point, $t^i\in\mathcal{M}$ we can ask how the
conformal data behaves under small variations. If a Lagrangian description is
available, this amounts to consider the deformation
\begin{equation} \label{eq:perturbation}
		\mathcal{L}_{t+\delta t} = \mathcal{L}_{t} + \delta t^i \mathcal{O}_i\,,
\end{equation}
and a standard perturbation analysis can be performed. More generally, we need
not make any special assumptions about the reference theory---in particular, it
is not necessary to assume deformations around a Gaussian fixed point produced
from a free theory. We can also consider non-Lagrangian cases by promoting
coupling constants to background sources and work directly with the effective
action $W[t^i(x)]$.

In a $d$-dimensional CFT, given a set of exactly marginal operators
$\mathcal{O}_i$ and their associated coupling constants $t^i$, the conformal
manifold is endowed with the Zamolodchikov metric, defined as the coefficient
of the two-point function of the marginal operators:\footnote{More
precisely, the marginal operators $\mathcal{O}_i$ give a basis of the conformal
manifold tangent bundle, and $\chi_{ij}$ are the components of the metric in
such a basis. Notice that this might not be a coordinate basis. For instance,
it can be taken to be an orthonormal basis in which $\chi_{ij}=\delta_{ij}$.}
\begin{equation}
		\left<\mathcal{O}_i(x_1)\mathcal{O}_j(x_2)\right>_t = \frac{\chi_{ij}(t)}{|x_{12}|^{2d}}\,,\qquad
		\Delta_i(t) = d\,.
\end{equation}
where $\left<\cdots\right>_t$ indicates that the correlator is calculated at a
given point of the conformal manifold. We stress that by marginal operator, we
mean an operator whose conformal dimension is strictly fixed to that of
spacetime \emph{throughout the conformal manifold} and never becomes
(ir)relevant at any point of $\mathcal{M}$, regardless of the mechanism
protecting it against quantum corrections.

By unitarity, $\chi_{ij}$ has the properties required to be a \emph{bona fide}
metric, and the Zamolodchikov metric allows us to define distances between two
points along a trajectory $\Gamma\subset\mathcal{M}$:
\begin{equation}
		\text{dist}_\chi = \int_\Gamma ds\,,\qquad ds^2 = \chi_{ij}\,dt^idt^j\,.
\end{equation}
This is the distance used in the CFT Distance Conjecture. Denoting the
parameter of a trajectory by $t$ with a slight abuse of notation, the
coordinates are given by $t^i(t)$, and under small variations we have:
\begin{equation}\label{trajectory-parametrization}
    \delta t^i \,\mathcal{O}_i = \delta t \, \mathcal{O} \, , \quad \mathcal{O} = \frac{dt^i}{dt} \, \mathcal{O}_i \, .
\end{equation}
This makes clear that each trajectory is related to a specific combination of
marginal operators, understood as a smooth vector field in the tangent bundle
of the conformal manifold. Reversing the logic, a result that is valid for 
any marginal operator holds for any trajectory in the conformal manifold. This is relevant for our
analysis, as we aim to obtain arguments that do not depend on the details of a
given family of CFTs and that are independent of the trajectory that we consider. 
Furthermore, from the previous equation we see that a
reparametrization of the trajectory amounts to a change in the normalization of
$\mathcal{O}$. In particular, this means that if $\mathcal{O}$ is chosen to be
properly normalized, then $t$ is nothing but the distance traveled along the
trajectory in terms of the Zamolodchikov metric.

As previously alluded to, we are interested in the conformal dimensions of
spin-$\ell$ operators and their behavior along a trajectory in the conformal
manifold as a function of the distance. In other words, we want to find the
conformal dimension as a function of $t$: $\Delta_\ell=\Delta_\ell(t)$. This
dependence is encoded in the two-point function $\left<J_\ell J_\ell\right>$.
Given a reference point, at first order in perturbation theory the variation
induced by a change $\delta t$ on this correlator is given by:
\begin{equation}\label{eq:first-order-CPT}
		\left\langle J_\ell(x_1) J_\ell(x_2) \right\rangle_{t+\delta t} = 
		\left\langle J_\ell(x_1) J_\ell(x_2) \right\rangle_{t} +
		\delta t \left. \int d^{d}y 	\left\langle J_\ell(x_1) J_\ell(x_2) \mathcal{O}(y) \right\rangle_{t} \right|_{\text{reg.}} \, .
\end{equation}
In a Lagrangian theory, equation \eqref{eq:first-order-CPT} can be understood
from the path integral formulation, but can also be obtained directly from a
variation of the effective action $W[t^i(x)]$. Note that the integral must be
regularized to take care of the divergences appearing as $y\to x_1 , x_2$,
which induces a renormalization of the operators $J_\ell$ to obtain a
well-defined quantity. This procedure is well defined, see e.g. reference
\cite{Komargodski:2016auf}, and can also be understood holographically
\cite{Berenstein:2014cia, Berenstein:2016avf}.

Up to this point we have only used perturbation theory, but not that the
deformation is exactly marginal and therefore that the resulting two-point
functions remain constrained by conformal invariance. The variation of the
correlator must thus take the generic form:
\begin{equation} \label{eq:ansatz}
		\left\langle J_\ell(x_1) J_\ell(x_2) \right\rangle_{t+\delta t} -	\left\langle J_\ell(x_1) J_\ell(x_2) \right\rangle_{t} = - \frac{H(x_{12})^{\ell}}{|x_{12}|^{2\tau}} \, 2 \, \delta \Delta_{\ell} \log |x_{12}| \,,
\end{equation}
with $\delta\Delta_\ell$ the change in the conformal dimension induced by
$\delta t$.\footnote{In general there could also be another term related to the
		change of the prefactor in front of the two-point function, i.e., the
		normalization of the spin $\ell$ operator. We however take this
operator to be in an orthonormal basis at any point in the conformal manifold.
This corresponds to a choice of renormalization scheme, and will not influence
out computations at the order we consider, see e.g. reference
\cite{Balthazar:2022hzb} for a detailed discussion.} This shift is therefore
encoded in the logarithmic divergence obtained after integration of the
three-point function given in equation \eqref{eq:first-order-CPT}. 

The integration over the three-point function in terms of the conformal
structure of type-$(\ell,\ell,0)$ correlators defined in equation
\eqref{conformal-structure-expansion} must be consistent with
\eqref{eq:first-order-CPT}, and must lead to the same type of divergences. This
means that the shift of $\Delta_\ell$ is controlled by a certain linear
combination of the OPE coefficients, $C^{\,}_{JJ\mathcal{O}}$:
\begin{equation}\label{variation-Delta}
	\delta \Delta_{\ell} = - \delta t\, C_{JJ\mathcal{O}} + \mathcal{O}(\delta t^{2})\,,\qquad
	C_{JJ\mathcal{O}} = \sum_{n=0}^\ell w_{n} C_{JJ\mathcal{O}}^{\,n} \, .
\end{equation}
This was shown in reference \cite{Sen:2017gfr}, where the renormalization
procedure was carefully performed by combining the embedding formalism with
dimensional regularization. The coefficients $w_{n}$---which we recall here for
completeness---were found to be given by: 
\begin{equation}\label{sen-tachikawa-coeffs}
    w_{n} = 2 \pi^{\frac{d}{2}} \, \frac{\Gamma(n+1)}{\Gamma(n+\frac{d}{2})} \, .
\end{equation}
Their precise values will not be relevant to us beyond the fact that, at first
order in conformal perturbation theory, the change in $\Delta_\ell$ is encoded
in a particular linear combination of the type-$(\ell,\ell,0)$ OPE
coefficients, which we refer to as the OPE combination $C_{JJ\mathcal{O}}$. 

Importantly, equation \eqref{variation-Delta} is true at any point in the
conformal manifold, which allows one to exponentiate it to its differential
version \cite{Behan:2017mwi}. This gives us an evolution equation for the
conformal dimensions in the conformal manifold as we change $t$:
\begin{equation} \label{eq:master-eq}
	\frac{d\Delta_{\ell}}{dt} = - \, C_{JJ\mathcal{O}}(t) \, .
\end{equation}
A possible cross-check for the above equation is that it transforms covariantly
under reparameterization of the marginal coupling $t$ discussed in equation
\eqref{trajectory-parametrization}, and consequently takes into account changes
in the normalization for the marginal operator $\mathcal{O}$.

As commented in Section \ref{sec:spinning-correlators}, the parity-odd
structures that can appear in $d=3$ do not contribute to equation
\eqref{eq:master-eq}. After integrating over the position of the marginal
operator, the leftover function of $x_1$ and $x_2$ is again parity odd. This is
inconsistent with \eqref{eq:ansatz}, which is parity even, and they therefore
cannot contribute to the variation of $\Delta_\ell$.

Knowing the OPE combination $C_{JJ\mathcal{O}}(t)$, we can thus in principle
reconstruct $\Delta_\ell$ at every point of the trajectory specified by
$\mathcal{O}$ via the evolution equation \eqref{eq:master-eq}, and measure the
distance traveled with respect to the Zamolodchikov metric. However, this
information of course is not readily available. Usually, one may try to extract 
information about $C_{JJ\mathcal{O}}(t)$ by also using conformal perturbation theory.
For instance, at first order the evolution of the three-point function 
$\left\langle J_\ell J_\ell \mathcal{O} \right\rangle$ is encoded into the four-point 
function $\left\langle J_\ell J_\ell \mathcal{O} \mathcal{O}\right\rangle$, 
leading to a similar differential equation for $C_{JJ\mathcal{O}}(t)$ 
\cite{Bashmakov:2017rko,Behan:2017mwi,Balthazar:2022hzb}. However, due to presence of this four-point
function, after imposing the usual associativity constraints, one is required
to have knowledge about the conformal data of operators appearing in the
various OPEs involving $\mathcal{O}$ and $J_\ell$, substantially complicating
the analysis. Furthermore, when applying this approach one should worry about the 
radius of convergence of conformal perturbation theory, since it amounts to reconstruct 
information of the whole conformal manifold starting from a single point by using 
conformal perturbation theory. Here we take a different approach, one more suited to
distinguishing between finite and infinite distance points in the conformal
manifold.

\subsection{A Criterion for Finite Versus Infinite Distance} \label{sec:finite-vs-infinite}

\begin{figure}[t]
\centering
	{\resizebox{0.7\textwidth}{!}{\includegraphics{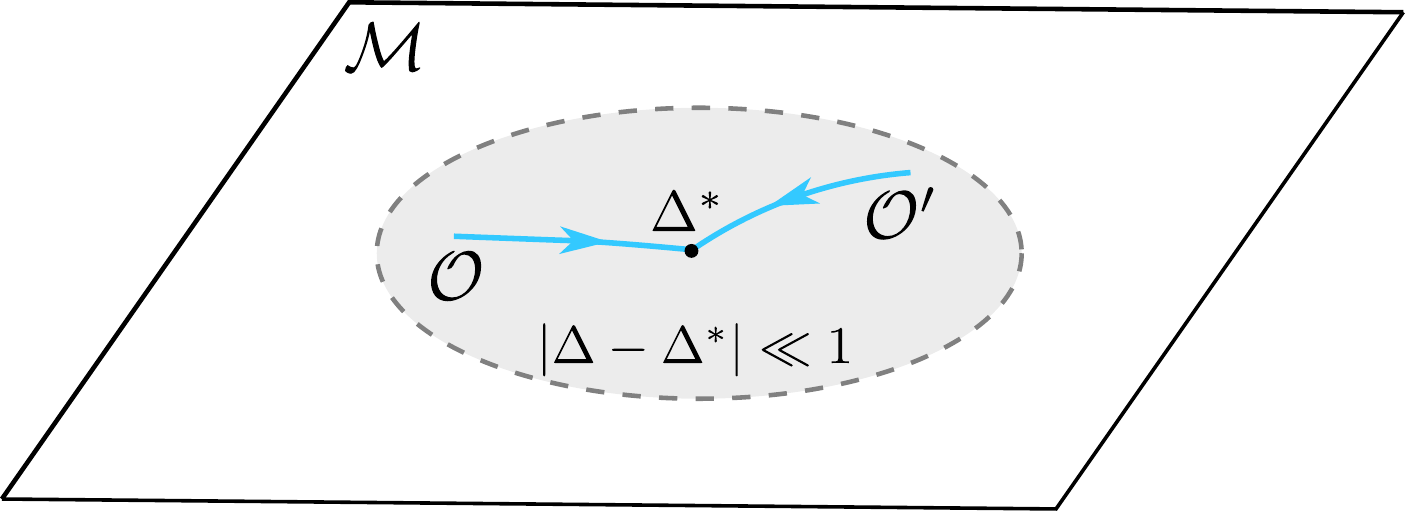}}}
	\caption{Different trajectories, corresponding to distinct marginal
	operators $\mathcal{O}$ and $\mathcal{O}^\prime$, in a neighborhood of the
	conformal manifold $\mathcal{M}$ leading to the same point with conformal
	dimension $\Delta^\ast$.}
    \label{fig:trajectories2}
\end{figure}

From here on, we will focus our attention on the neighborhood of a point where
certain spin-$\ell$ operators have dimension $\Delta_\ell^\ast$---or similarly
for a locus defined by $\{\Delta_\ell(t) = \Delta_\ell^\ast\}$. As in the
previous section, we also consider a trajectory specified by a marginal
operator $\mathcal{O}$, understood as a smooth vector field in the neighborhood
of the point under consideration. One can think of different choices of
$\mathcal{O}$ as different ways of approaching this point, as depicted in
Figure \ref{fig:trajectories2}. In particular, drawing conclusions about 
any marginal operator in this neighborhood is equivalent to finding properties 
of any trajectory approaching this point in any possible direction.

The idea behind our criterion to characterize whether a point is at (in)finite
distance from a generic point in the interior of $\mathcal{M}$ is then the
following: if one is able to obtain $C_{JJ\mathcal{O}}$ as a function of
$\Delta_{\ell}$ close to the point under consideration, one can recover the
behavior of the conformal dimension as a function of $t$ in that neighborhood
using the evolution equation \eqref{eq:master-eq}. As argued above, when the
two-point function of $\mathcal{O}$ is chosen in an orthonormal basis, $t$ is
then interpreted as the Zamolodchikov distance traveled along the trajectory
specified by $\mathcal{O}$ itself. We can then use the behavior of
$\Delta_{\ell}(t)$ as it approaches $\Delta_{\ell}^{\ast}$ to learn whether
this point is at finite or infinite distance along the given trajectory.

Notice however that one can only obtain non-trivial information about the
traveled distance when the conformal dimension varies along the trajectory. We
hence need to pick a spin-$\ell$ operator whose conformal dimension is not
fixed to a specific value over the whole conformal manifold. For instance, the
energy-momentum tensor, conserved flavor currents, and protected BPS operators
in supersymmetric theories have this property, and are therefore excluded from
this analysis. We will thus assume that the higher-spin operators are conserved
only at specific points or loci of the conformal manifold.

\medskip

To illustrate this idea let us start with the most trivial---but also most
generic--- example, namely the case where the OPE combination can be considered
constant in the neighborhood defined by $\Delta^\ast_\ell$:
$C_{JJ\mathcal{O}}\to C_{JJ\mathcal{O}}^{*}\neq0$ as
$\Delta_{\ell}\to\Delta_{\ell}^{*}$. Defining the change in conformal dimension
as $\gamma_\ell$,\footnote{By abuse of notation, in this section we use
$\gamma_\ell = \Delta_\ell - \Delta_\ell^\ast$ for the deviation of the
conformal dimension with respect to an arbitrary point of the conformal
manifold in analogy to the usual anomalous dimension $\gamma = \Delta -
\Delta_\text{free}$. When the reference point corresponds to a free theory,
$\Delta^\ast = \Delta_\text{free}$, as will be the case in the next sections,
the two quantities coincide.} the evolution equation \eqref{eq:master-eq}
reduces to:
\begin{equation}
		\frac{d\gamma_{\ell}}{dt} = - \, C_{JJ\mathcal{O}}^{\,*} + \mathcal{O}(\gamma_\ell)\,,\qquad C_{JJ\mathcal{O}}^{\,*}=\text{const}\,,\qquad
		\gamma_\ell = \Delta_\ell -\Delta_\ell^\ast \, .
\end{equation}
Therefore, the leading behavior in the change of conformal dimension is given
simply as a linear approximation
\begin{equation} \label{eq:solution}
		\gamma_{\ell}(t) \simeq \gamma_{\ell}^{0} - C_{JJ\mathcal{O}}^{\,*} \, t\, ,
\end{equation}
where we have fixed the integration constant in terms of the deviation at the
point where we put the origin of the trajectory, namely
${\gamma}_{\ell}(t=0)=\gamma_{\ell}^{0}$. Of course, this approximation is only
valid when the deviation $\gamma_{\ell}(t)$ is sufficiently small. 

We have therefore obtained the solution to equation \eqref{eq:master-eq} close
to ${\gamma}_{\ell}=0$ when $C_{JJ\mathcal{O}}$ takes a non-zero value at this
point. Importantly, the solution given in equation \eqref{eq:solution} reaches
$\gamma_{\ell}=0$ for finite values of $t$. In other words, this point is at
finite distance from the initial point $t=0$ along the trajectory defined by
$\mathcal{O}$. 

To summarize, we see that if at some point in the conformal manifold we have
$C_{JJ\mathcal{O}}^{*}\neq 0$ for a given marginal operator $\mathcal{O}$, then we
have found a trajectory reaching this point within finite distance.
Furthermore, finding a single trajectory for which this is true is enough to
imply that the point under consideration is a finite-distance point in the
conformal manifold.

\medskip

Let us now give an example for which the point is reached at infinite distance.
If the OPE coefficients grows linearly with the deviation,
\begin{equation} \label{eq:parametrization}
		C_{JJ\mathcal{O}}({\gamma}_\ell) \simeq \alpha_\chi \, {\gamma}_\ell\,,\qquad
		\gamma_\ell = \Delta_\ell-\Delta^\ast_\ell\,,
\end{equation}
for some constant $\alpha_\chi>0$, the evolution equation discussed above is
solved by an exponential. The behavior of the deviation as it reaches zero is
of the form:
\begin{equation}
		{\gamma}_\ell = {\gamma}_\ell^0\,e^{-\alpha_\chi \, t}\,,
\end{equation}
with the integration constant again fixed by
${\gamma}_{\ell}(t=0)={\gamma}_{\ell}^{0}$. As $t$ is again interpreted as the
distance in the conformal manifold, we have the case of a trajectory with the
exponential behavior typical of the CFT Distance Conjecture, see equation
\eqref{exponential-SDC}. Note that ${\gamma}_\ell\to0$ for $t\to\infty$
regardless of the point where we place $t=0$, which means that this point is at
infinite distance along the trajectory defined by the marginal operator
$\mathcal{O}$. However, this does not mean that we have an infinite-distance
point. Indeed, it could be that $\mathcal{O}$ is associated with a
highly-non-geodesic trajectory, but there exists another trajectory for which
the point can be reached within a finite distance.

\medskip

To characterize whether a point where the conformal dimension is given by
$\Delta_\ell^\ast$ is reached at finite distance or not along certain
trajectory, let us consider a general case where $C_{JJ\mathcal{O}}$ is
parameterized by:\footnote{This parametrization is motivated by a perturbative
description valid around the point under consideration, in which both
$C_{JJ\mathcal{O}}$ and ${\gamma}_\ell$ enjoy an expansion in powers of a small
parameter. For instance, this will be the case when we discuss HS points in
later sections.}
\begin{equation}\label{eq:power-case}
	C_{JJ\mathcal{O}}({\gamma}_{\ell}) \sim {\gamma}_{\ell}^{1+a} \qquad \text{ as } {\gamma}_\ell=\Delta_\ell-\Delta^\ast_\ell\to0 \, .
\end{equation}
Close to the reference point, and up to subleading corrections, the deviation
from $\Delta_\ell^*$ then behaves as:
\begin{equation}
		t \sim\,
		\left\{\,\begin{aligned}
			\log\, {\gamma}_\ell\,,\quad&\text{if } a=0 \, ,\\
			{\gamma}_\ell^{-a}\,,\quad&\text{if } a\neq0 \, .\\
		\end{aligned}\right.
\end{equation}

\begin{figure}[t!]
\centering
\begin{subfigure}[t]{.49\linewidth}
		\centering
		{\resizebox{0.5\textwidth}{!}{\includegraphics{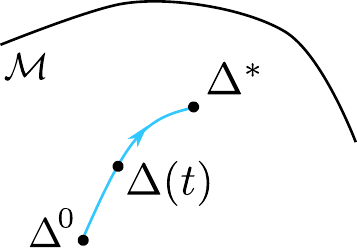}}}
		\vspace*{3mm}
		\caption{Finite distance}
	\end{subfigure}
\hfill\begin{subfigure}[t]{.49\linewidth}
		\centering
		{\resizebox{0.5\textwidth}{!}{\includegraphics{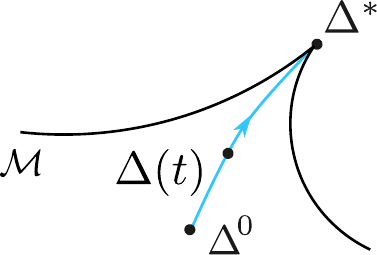}}}
		\vspace*{3mm}
		\caption{Infinite distance}
	\end{subfigure}
\caption{Two different trajectories associated with a marginal operator in the
		neighborhood of a point with dimension $\Delta^*$. If the OPE
		combination $C_{JJ\mathcal{O}}(t)$ remains approximately constant in
		the neighbourhood of the point, it is a finite-distance point. On the other
		hand, if $C_{JJ\mathcal{O}} \sim (\Delta-\Delta^*)^{1+a}$ with
		$a\geq0$, it is at infinite distance from any other point.
  \label{fig2}
}
\end{figure}

For the same reasons as above, $t$ measures the distance from the distance from
an arbitrary point encoded in the integration constant. The case $a<0$
corresponds to a finite distance, while $a\geq0$ is at infinite distance. For
$a=0$ we have not only the critical case between finite and infinite distance,
but also the one where the deviation ${\gamma}_\ell$ decreases exponentially
fast with the distance, that is the one associated with the third part of the
CFT Distance Conjecture explained in the introduction. This is depicted in
Figure \ref{fig2}.

Once again, we stress that if $a\geq0$, this does not mean that the point at
$t=0$ is an infinite-distance point, but could be associated with a
non-geodesic trajectory and reach the point in an seemingly infinite-distance
fashion. In order to ensure that it is truly an infinite-distance point, we
need to show that it is the case for \emph{any} trajectory. Conversely, a
marginal operator inducing a trajectory where a point is at finite distance is
sufficient to imply it is a finite-distance point, albeit it is maybe not
attained via the shortest path. We have therefore found the following
criterion.

\paragraph{CFT Distance Criterion:} In a neighborhood of a point of a conformal
manifold $\mathcal{M}$ where higher-spin operators have conformal dimension
$\Delta_\ell^\ast$, the behavior of the OPE combination $C_{JJ\mathcal{O}}$
defined in equation \eqref{variation-Delta} as
$\gamma_\ell=\Delta_\ell-\Delta_\ell^\ast\to0$ is sufficient to decide whether
it is a:\footnote{Even though the criterion is tailored to the behavior given
in equation \eqref{eq:power-case} close to the reference point, a similar
analysis can be performed for other parametrization.}
\begin{itemize}
		\item[-] Finite-distance point:\qquad $\exists\,\mathcal{O} \text{ such that } 
			C_{JJ\mathcal{O}} \sim {\gamma}_{\ell}^{1+a} \text{ with } a<0\,,$ 
		\item[-] Infinite-distance point:\qquad $C_{JJ\mathcal{O}} \sim {\gamma}_{\ell}^{1+a}
			\text{ with } a\geq 0 \ \forall \mathcal{O}\,.$
\end{itemize}
\medskip

Note that this criterion does not depend on the precise nature of the operator
$J_\ell$, and at no point did we assume a specific value of $\ell$, and
should be valid for an conformal primary. To apply it, we only need an
operator whose conformal dimension is not constant along the whole trajectory.
Therefore, any operator for which this reasoning is applicable should give us
the same answer. In particular, there cannot be an operator placing a point at
finite distance while another predicts an infinite-distance behavior. This
implies non-trivial relations between three-point functions of the form
$\langle J_\ell J_\ell\mathcal{O}\rangle$ and the possible marginal operators
for the conformal manifold to make sense as a metric space.
 
In this section, we have discussed how to differentiate between finite- and
infinite-distance points of the conformal manifold, which typically requires
understanding the behavior of $C_{JJ\mathcal{O}}$ close to that point for one
or potentially even all marginal operators of the theory, an information that
may be challenging to access in general. Despite this hurdle, we can draw an
interesting observation: any point where $C_{JJ\mathcal{O}}$ is non-vanishing
for any marginal operator is a finite-distance point. Conversely, a vanishing
OPE combination is a necessary, albeit not sufficient, condition to have an
infinite-distance point.

\section{Higher-Spin Points are at Infinite Distance}\label{sec:HS-at-infinite-distance}

Having a criterion to decide whether a point of the conformal manifold
$\mathcal{M}$ is at infinite distance, we now apply the procedure described in
the previous section to higher-spin (HS) points. There, (part) of the HS
operators become conserved currents and the conformal group enhances to a
larger, infinite-dimensional higher-spin symmetry.

In order to apply the CFT Distance Criterion to show that HS points are at
infinite distance, we will assume that a given HS operator is never conserved
everywhere in the conformal manifold, as this can happen only if it is part of
a free subsector that is completely decoupled from the rest of the theory at
all points of $\mathcal{M}$. Moreover, following the discussion at the end of
the previous section, it is in principle enough to show that HS points are at
infinite distance for an operator with e.g. spin $\ell=4$; there is indeed
always a spin-four conserved current at any HS point \cite{Maldacena:2011jn,
Stanev:2013qra, Boulanger:2013zza, Alba:2013yda, Alba:2015upa}. While we could
in principle set $\ell=4$ from now on, we will keep the spin arbitrary and
specialize to the spin-four case for illustrative purposes.

\subsection{Conservation Laws and Higher-Spin Correlators}\label{sec:charge-conservation}

At the HS point, the anomalous dimension of the higher-spin operator vanish and
its conformal dimension saturates the unitarity bound for spin-$\ell$
operators: $\Delta_{\ell}=d-2+\ell$. As is well known in Conformal Field
Theory, when this happens the associated conformal multiplet develops a null
state corresponding to a conservation condition:
\begin{equation}\label{Ward-identity}
		\partial\cdot J_\ell = \partial^{\mu} J_{\ell,\,\mu\nu_1\dots\mu_\ell}\,\xi^{\nu_2}\dots\xi^{\nu_\ell}=0 \, .
\end{equation}

This identity can be used to constrain correlators involving these conserved
currents, as their divergences have to vanish up to contact terms. We will
often refer to this as the \emph{current-conservation condition}, and it will
allow us to reduce the number of conformal structures that can appear in a
given correlator. Starting with general type-$(\ell_1,\ell_2,\ell_3)$
correlators and taking the divergence with respect to a HS operator, the
current conservation conditions will impose constraints on its OPE
coefficients. An important observation is that the null states associated with
the divergence of conserved currents transform as spin-$(\ell-1)$ conformal
primaries, see e.g. reference \cite{Costa:2011mg}. The current-conservation condition of
a type-$(\ell_1,\ell_2,\ell_3)$ correlator with respect to the first current is
therefore equivalent to the vanishing of a type-$(\ell_1-1,\ell_2,\ell_3)$
three-point function. As was already realized in reference
\cite{Maldacena:2011jn}, since both the type-$(\ell_1,\ell_2,\ell_3)$ and
type-$(\ell_1-1,\ell_2,\ell_3)$ correlators can be written in terms of the same
building blocks, see Section \ref{sec:spinning-correlators}, we can instead
write the divergence as a differential operator acting on the space of these
building blocks. This considerably simplifies an otherwise rather cumbersome
computation. The precise form of this operator can be found in reference
\cite{Zhiboedov:2012bm}, and we use it for cases of interest in Appendix
\ref{sec:appendix-A}.

As already mentioned in Section \ref{sec:spinning-correlators}, we note that
parity-even and -odd structures do not mix in the current-conservation
condition, as divergences preserve parity. To verify this, one can observe that
the spin changes by one when taking the divergence, and that spin-$\ell$
operators acquires a sign $(-1)^\ell$ under parity. This---together with the
fact that parity-odd structures do not contribute to equation
\eqref{eq:master-eq}---allows us to ignore all allowed parity-odd
$(d=3)$-structures for our purposes.

\medskip

By Noether's theorem, the HS conserved currents also have associated conserved
charges by integrating them over a codimension-one surface $\Sigma$. Note that,
contrary to the usual flavor charges, higher-spin charges have spin and we
again use polarization vectors to contract the extra-indices. We then obtain a
formal one-form that can be used to construct the charge:\footnote{ A more
general construction involves contracting the $(\ell-1)$ remaining indices with
a conformal Killing tensor $\zeta$ satisfying
$\partial^{(\mu}\zeta^{\nu_1\dots\nu_{\ell-1})}=0$. For our purpose however, it
will be sufficient to consider constant polarization vectors, which forms a
subclass of those tensors. We defer to reference \cite{Boulanger:2013zza} for a
review of the construction of the space of conformal Killing tensors and the
associated charges.}
\begin{equation} \label{charges}
		Q_{\ell}(\xi) = \int_\Sigma \check{J}_\ell(\xi)\,,\qquad \check{J}_\ell(\xi) =
		J_{\ell,\,\mu\nu_1\dots\nu_{\ell-1}}\,\xi^{\nu_1}\dots\xi^{\nu_{\ell-1}}dx^\mu\,.
\end{equation}
While the textbook approach to conserved charges involves integrating over the
$(d-1)$ space coordinates, the usefulness of more general surfaces and induced
topological properties of the charges have long been recognized, and have
recently come in full force in the context of higher-form symmetries
\cite{Gaiotto:2014kfa}. Note however that the charges $Q_\ell(\Sigma,\xi)$ have
spin $\ell-1$ and act on (possibly-spinning) local operators and the currents
transform in $\ell$-symmetric-traceless representations of the Lorentz group.
While they do generalize the notion of ordinary spin-one flavor charges, they
do not correspond to the type of generalized symmetries discussed in reference
\cite{Gaiotto:2014kfa}, which act on extended objects and the involved currents
transform in \emph{anti}-symmetric spacetime representations.

The conserved charges can be also used to constrain correlators involving HS
conserved currents. Indeed, when inserted in a correlator, we must take into
account contact terms---that is, contributions at coincident points:
\begin{equation}\label{ward-contact-terms}
		\left<(\partial\cdot J_\ell)(y,\xi)\Phi_1(x_1)\cdots\Phi_n(x_n)\right> \propto \sum_{k=1}^n\delta(y-x_k)\left<\Phi_1(x_1)\cdots[Q(\xi),\Phi_k(x_k)]\cdots\Phi_n(x_n)\right>\,,
\end{equation}
up to a constant related to normalization, and the operators $\Phi_n$ are
possibly spinning. Integrating over a $d$-dimensional submanifold of spacetime
containing only a single position $x_k$, we can therefore select a given
commutator and reduce the number of operators appearing in the correlator. On
the other hand, using Stokes' theorem on the LHS of equation
\eqref{ward-contact-terms}, we can integrate the current, understood as a
formal one-form as in equation \eqref{charges}, over the codimension-one
surface $\Sigma$ bounding the submanifold. The charges are of course
topological and we can deform the surface as long we are not crossing over
another point, in which case we must take into account the appropriate
commutator. We will refer to this as the \emph{integrated Ward identity} and it
will allow us to relate the OPE coefficients in a given correlator through
properties of the HS algebra we review below.

\medskip

Given the conserved charges, there is a further set of identities that one can
impose on correlators to constrain not only their OPE coefficients but also how
the charges can act on different operators. These are particularly interesting,
since they allow us to impose the coexistence of HS symmetry with other
ingredients of the theory like, e.g. the presence of the energy-momentum
tensor. In fact, these constraints were crucial for the original derivation of
the theorems relating HS points to free theories \cite{Maldacena:2011jn,
Stanev:2013qra, Boulanger:2013zza, Alba:2013yda,Alba:2015upa}. Following
reference \cite{Maldacena:2011jn}, we will call them \emph{charge-conservation
identities}. Given a set of operators, the idea is to consider the correlator
of their commutator with $Q_\ell$. Since the conserved charge must annihilate
the vacuum, this correlator must vanish. This is:
\begin{equation} \label{general-charge-conservation}
    0 = \left<[Q_\ell(\xi),\Phi_1(x_1)\cdots\Phi(x_n)]\right> = \sum_{k=1}^n \left<\Phi_1(x_1)\cdots[Q_\ell(\xi),\Phi_k(x_k)]\cdots\Phi_n(x_n)\right> \, .
\end{equation}
This can also be obtained from equation \eqref{ward-contact-terms} by
integrating over the entire of spacetime or any other submanifold that does not
have a boundary, and containing all the positions $x_k$. Using Stokes'
theorem, the result must vanish.

Using the HS algebra to further expand each commutator in equation
\eqref{general-charge-conservation} as a sum over operators appearing in the
various HS multiplets, one can infer a set of constraints involving both the
parameters controlling the action of $Q_\ell$ on the operators and their OPE
coefficients. In what follows, we review some of the properties of the HS
algebra, give an example of such a charge-conservation identity, and its
consequences.

\medskip

The commutation relations of the charges must close and generate the full
higher-spin algebra. In general, the action of the charges on an operator of
twist $\tau=\Delta-\ell$ will involve derivatives of the various fields falling
inside the same higher-spin multiplet so that the total twist is conserved.
Tracking the index structure of such commutation relations can however be
cumbersome, and it is instead very convenient to focus on a particular
spacetime direction in order to simplify the index structure. Following
reference \cite{Maldacena:2011jn}, we will work in light-cone coordinates:
\begin{equation}
		ds^2=dx^+dx^- + \delta_{ij} dx^idx^j\,, \qquad i,j=2,\dots,d-1\,,
\end{equation}
and use polarization vectors forcing the indices to be in the
$(x^-)$-direction: $\xi^\mu =\delta^\mu_-$.\footnote{In terms of group theory,
this corresponds to consider the celebrated non-compact $\mathfrak{sl}_2$
derivative subsector, where the higher-weight states are the
$\ell$-traceless-symmetric conformal primaries. The only relevant generator
actions are then associated with derivatives of the fields in the
$x^-$direction.} To differentiate HS operators in this regime from those with
arbitrary polarizations, written with uppercase letters, e.g. $J_\ell(x,\xi)$,
we will use lowercase letters to indicate spinning operators whose indices all
point in that direction. For instance:
\begin{equation}
		j_\ell = J_{\ell,-\dots-}\,.
\end{equation}
The charges are then defined in terms of the currents with the indices in that
direction
\begin{equation}\label{charges-minus}
		Q_{\ell} = \int_\Sigma dx^{-} d^{d-2}x \;\; j_{\ell}(x) \,,
\end{equation}
where now $\Sigma$ is taken as a ``slab'' of spacetime in the directions
perpendicular to $x^+$. For simplicity, we will henceforth drop the dependence
on the polarization of the charges whenever the context is clear:
$Q_\ell(\delta^\mu_-)=Q_\ell$. 

First, since the charges have twist zero, each term appearing in commutators
involving $Q_\ell$ and an operator $k_{\ell'}$ of twist $\tau$ must also have
twist $\tau$. Moreover, these can only involve twist-$\tau$ operators
$k_{\ell''}$ with all indices in the minus direction as well as
$x^-$-derivatives, $\partial=\frac{\partial}{\partial x^-}$. Due to this index
structure, each derivative increases the spin by one but the twist remains
unchanged. The most general formula for such a commutator is then:
\begin{equation} \label{Qkcomm}
		[Q_\ell, k_{\ell'}] = \sum_{\ell''=\text{max}(\ell'-\ell+1,0)}^{\ell+\ell'-1} \eta_{\ell''}\, \partial^{\ell+\ell'-1-\ell''} k_{\ell''}\,.\\
\end{equation}

As discussed above, the coefficients $\eta_\ell$ can be constrained---together
with correlators---using the charge-conservation identities, and lead to a
number of useful relations \cite{Maldacena:2011jn, Stanev:2013qra,
Boulanger:2013zza, Alba:2013yda, Alba:2015upa}. The simplest example is
obtained by acting with $Q_\ell$ on the two-point function
$\left<k_{\ell'}k_{\ell''}\right>$. From this, one can show that if
$k_{\ell''}$ appears in $[Q_\ell, k_{\ell'}]$ then $k_{\ell'}$ must appear in
$[Q_\ell, k_{\ell''}]$.\footnote{One might worry that, for $\tau \geq d$, there may be extra terms in the commutator in equation \eqref{Qkcomm} of the form $\Box^{n} \mathcal{V}_{\ell''}$ with $\mathcal{V}_{\ell''}$ a twist-$(\tau-2n)$ operator. However, through a similar argument applied to the correlator $\left<k_{\ell'}\mathcal{V}_{\ell''}\right>$, one can show that such terms cannot appear.}

Moreover, since we consider local theories, there is a spin-two
conserved current that generates conformal transformations, the energy-momentum tensor $T_{\mu\nu}$.
Additional constraints can be obtained by specifically considering
charge-conservation identities on correlators involving $t_2 = T_{--}$, as they
enable us to further constrain the appearance of a given operator in
commutators given in equation \eqref{Qkcomm}. For instance, since $t_2$ is
related to the dilatation operator and charges have conformal dimension
$\ell-1$, $j_\ell$ must always appear in $[Q_\ell,t_2]$. Then, using this fact
and charge-conservation identities, one can further show that not all \emph{a
priori} possible conserved currents appear in $[Q_\ell,t_2]$. When $\ell=4$,
one finds that the most general form of the commutator is given by:
\begin{equation}\label{Q4j2}
		[Q_4, t_2] = \alpha_0\, \partial^5 j_0 +\alpha_1\,\partial^4j_1+ \alpha_2\, \partial^3 j_2 + \alpha_3\,\partial^2 j_3 + \alpha_4\,\partial j_4\,.
\end{equation}

By the argument above, $j_4$ must appear in the
commutator and thus $\alpha_4\neq0$.
Notice that we do not need to assume that the energy-momentum tensor $t_2$ is the unique spin-two conserved current. This is, $j_2$ is a \emph{generic} spin-two conserved current, that can include contributions from both $t_2$, and possibly any additional spin-two conserved current.
When $t_2$ is the unique spin-two conserved current, not all coefficients $\alpha_\ell$ with
$\ell=0,\cdots,5$ are non-zero, and it is possible to show that $\alpha_1=0=\alpha_3$ \cite{Maldacena:2011jn}.

Restricting the possible operators appearing
the various commutators will prove very powerful. It will allow us to show that
some of the three-point functions involving marginal operators must vanish, and
consequently will be crucial to prove that HS point are at infinite distance.

\subsection{Higher-Spin Conserved Currents and Marginal Operators}\label{HS-conserved-currents-condition}

Following the CFT Distance Criterion discussed in Section
\ref{sec:finite-vs-infinite}, a necessary condition for a HS point to be at
infinite distance is that the OPE combination must vanish for any marginal
operator at this point: $C_{JJ\mathcal{O}}^\text{HS}=0$. We will thus now shift
our focus to three-point functions $\left< J_\ell J_\ell\mathcal{O}\right>$.
Our goal is therefore to show that these correlators are always trivial,
regardless of the chosen marginal operator.

To do so, we will first use the current-conservation condition on this
correlator. This will impose constraints on the various OPE coefficients,
reducing the number of allowed conformal structure to a single one. By further
using an integrated Ward identity, we will then show that the remaining \emph{a
priori} independent parameter must vanish.

\medskip

Let us start by imposing conservation of the higher-spin currents, see equation
\eqref{Ward-identity}, on this three-point function. Up to contact terms, we
have:
\begin{equation} \label{eq:divergenceless-condition}
	\partial_2 \cdot \left< J_{\ell}(x_1) J_{\ell}(x_2) \mathcal{O}(x_3)\right> = 
	\left< J_{\ell} (x_1) (\partial\cdot J_{\ell})(x_2) \mathcal{O}(x_3)\right> = 0 \,,
\end{equation}
where $\partial_i$ is understood as taking the divergence with respect to the
operator located at $x_i$. As discussed in Section
\ref{sec:charge-conservation}, the LHS can be conveniently evaluated using a
differential operator in the space of building blocks $H$, $V_1$, and $V_2$.
Its expression, as well as the result for this current-conservation condition,
can be found in Appendix \ref{sec:appendix-A}. Here, we will simply give a
qualitative description of the procedure needed to show that at the higher-spin
point, this three-point function and the related OPE combination vanish.

As we have seen in Section \ref{sec:spinning-correlators}, for a fixed value of
$\ell$ there are $\ell+1$ independent type-$(\ell,\ell,0)$ conformal
structures, as defined in equation \eqref{conformal-structure}. Taking the
divergence as in equation \eqref{eq:divergenceless-condition} we then obtain a
linear combination of conformal structures of type $(\ell,\ell-1,0)$. Since
there are only $\ell$ such structures, demanding that the resulting three-point
function vanish leads to a set of $\ell$ linear constraints on the $\ell+1$ OPE
coefficients $C_{JJ\mathcal{O}}^{\,n}$. If these $\ell$ constraints are
independent, we can express all the OPE coefficients in terms of a single one,
and the solution takes the form:
\begin{equation} \label{eq:recursion-relation}
		C_{JJ\mathcal{O}}^{\,n} = v^{n} \, C_{JJ\mathcal{O}}^{\,\ell} \, ,\qquad n=0,\dots,\ell \,, 
\end{equation}
where the components of the vector, denoted as $v^n\in\mathbb{R}$, are normalized such that $v^\ell=1$.
This is found by showing that the current-conservation constraint in equation
\eqref{eq:divergenceless-condition} can be viewed as a recursion relation in
the space of OPE coefficients $C_{JJ\mathcal{O}}^{\,n}$. In general, solving
this for arbitrary values of $\ell$ is difficult, but not particularly
illuminating. It is on the other hand straightforward to do it on a
case-by-case basis. For instance, when $\ell=4$ the coefficients are given by:
\begin{equation}\label{vn-spin-4}
    v^n = \left( \frac{-24}{(d-2)d(d+2)(d+6)} \, ,\, \frac{48(d+3)}{(d-2)d(d+2)(d+6)} \, ,\, \frac{72}{(d-2)d(d+6)} \, ,\, \frac{-8(d+7)}{(d-2)(d+6)} \, ,\, 1 \right) \, .
\end{equation}
However, the specific values of $v^n$ will not be relevant for our analysis,
only that the current-conservation constraint reduces the number of allowed
structure from $(\ell+1)$ to only one. The details of this computation can be
found in Appendix \ref{sec:appendix-A}.

To see the impact of this constraint on the conformal-perturbation-theory
equations, let us write the OPE combination at the HS point after imposing the
Ward identity \eqref{eq:divergenceless-condition}:
\begin{equation} \label{eq:3pt-HS-point}
	C^{\text{HS}}_{JJ\mathcal{O}} = C_{JJ\mathcal{O}}^{\ell} \, \sum_{n=0}^{\ell} w_{n} v^{n} \, .
\end{equation}
While one could hope that $\sum_{n=0}^{\ell} w_{n} v^{n}=0$, directly showing
that $C_{JJ\mathcal{O}}^{\text{HS}}=0$, this is unfortunately generically not
the case. We have checked this explicitly for $4 \leq \ell \leq 10$ in any
dimension by using the expression of coefficients $w_{n}$, given in equation
\eqref{sen-tachikawa-coeffs}, and those of $v^n$, obtained by solving the
recursive relation obtained in Appendix \ref{sec:appendix-A}.\footnote{For
$\ell=2,$ and $\ell=3$, it turns out that this quantity does vanish for $d=4$
and $d=14$, respectively.} In fact, it is hard to believe that there is a given
spin $\ell$ for which it vanishes for any dimension. The strategy is then to
further show that the yet-undetermined OPE coefficient
$C_{JJ\mathcal{O}}^{\,\ell}$ has to vanish. Doing so will establish that
$C^{\text{HS}}_{JJ\mathcal{O}}=0$---as expected---in a way that does not depend
on the precise values of $v^n$ and $w_n$.

\medskip

To show that this is the case, we now turn to the integrated Ward identity,
which allows us to reduce the correlator to a two-point function. Fixing the
coordinates $x_2$ and $x_3$, we can use a version of the Ward identity
\eqref{ward-contact-terms} applied to our case and integrate the current over a
submanifold only containing $x_2$, and we obtain:
\begin{equation} \label{eq:Ward1}
		\int_{\Sigma} \left\langle \check{J}_{\ell}(x_1,\xi_1) J_{\ell}(x_2,\xi_2) \mathcal{O}(x_3) \right\rangle \propto \left\langle \left[Q_{\ell}(\xi_1), J_{\ell}(x_2,\xi_2)\right] \mathcal{O}(x_3) \right\rangle \, .
\end{equation}
The codimension one surface $\Sigma$ is the boundary of the submanifold
containing $x_2$. The integral is then performed with respect to the first HS
conserved current $\check{J}_\ell$, where the first index is in direction
normal to $\Sigma$, see equation \eqref{charges}.

As discussed above, the commutator of an operator with a conserved charge must
involve only operators of the same twist, and in this case they must therefore
all have $\tau=d-2$. However marginal operators have twist $\tau=d$, and as a
consequence there cannot be any operator appearing in the commutator so that
the two-point function on the RHS of equation \eqref{eq:Ward1} does not vanish,
and we conclude that:
\begin{equation}
  \left\langle \left[Q_{\ell} ,J_{\ell}(x_2)\right] \mathcal{O}(x_3) \right\rangle = 0 \, .
\end{equation}

By consistency, the integral on the LHS of equation \eqref{eq:Ward1} must also
vanish. If we choose $\Sigma$ such that it picks the component of $\check{J}$
in the direction of $\xi_1$, we can then expand the correlator in terms of the
conformal structures, and using equation \eqref{eq:recursion-relation}, we can
factor out the so-far-undetermined coefficient $C_{JJ\mathcal{O}}^{\,\ell}$ to
obtain:
\begin{equation}
		\int_{\Sigma} \left\langle J_{\ell}(x_1,\xi_1) J_{\ell}(x_2,\xi_2) \mathcal{O}(x_3) \right\rangle = C_{JJ\mathcal{O}}^{\,\ell} \int_{\Sigma} \left( \sum_{n=0}^{\ell} v^{n}\, \Theta_{n}(x_i;\xi_1,\xi_2) \right) \overset{!}{=} 0\, .
\end{equation}
If we can find polarization vectors $\xi_1,\xi_2$ so that the integral gives a
non-zero result, the constraint coming from the integrated Ward identity can
only be satisfied if $C_{JJ\mathcal{O}}^{\,\ell}=0$, as desired. In Appendix
\ref{app:surface-integrals}, we show explicitly that the integral indeed does
not vanish for a particular choice of polarization vectors. We therefore
conclude that at any HS point and for any marginal operator $\mathcal{O}$, the
three-point function $\left\langle J_{\ell} J_{\ell} \mathcal{O} \right\rangle$
is trivial, and by extension so is the OPE combination relevant to the CFT
Distance Criterion:
\begin{equation} \label{JJO vanishes}
    C^{\text{HS}}_{JJ\mathcal{O}}=0\,.
\end{equation}
This result is in the end not too surprising. In fact, it is known that when
the two spins are different, the OPE coefficients vanish, see e.g. references
\cite{Giombi:2011rz, Skvortsov:2015pea}. For equal spin however, the
conservation condition is not enough to come to this conclusion, as one must
use the integrated Ward identity.

For a theory of 4d free vectors, which include $\mathcal{N}=4$ super-Yang--Mills in the free-field limit, this result can be obtained from a direct
computation by decomposing the field-strength into (anti-)self dual components
$F=F_++F_-$. The only non-vanishing two-point function of the
field-strength is $\langle F_+F_-\rangle$, and the HS currents and marginal
operators can be written as \cite{Gelfond:2006be}:
\begin{equation}
		J_\ell\sim F_+\partial^\ell F_-\,,\qquad \mathcal{O}\sim F_+F_+ + F_-F_-\,,
\end{equation}
where we have been very cavalier with the index structure. By Wick's theorem, the relevant three-point functions then directly vanish:\footnote{We thank E. Skvortsov for pointing out this very concise argument to us.}
\begin{equation}
		\langle (F_+\partial^\ell F_-)\, (F_+\partial^\ell F_-)\, (F_+F_+)\rangle = 0 = 
		\langle (F_+\partial^\ell F_-) \, (F_+\partial^\ell F_-)\, (F_- F_-)\rangle \, .
\end{equation}

\medskip

As advertised, we have therefore found that the OPE combination always vanishes at any HS
point, $C^\text{HS}_{JJ\mathcal{O}}=0$. We stress that this result is not
sensible to the actual numerology of the coefficients $w_{n}$ and $v^n$, but is
rather a consequence of the properties of the higher-spin symmetry.

This result shows that higher-spin points are not in the regular
finite-distance class for which $C_{JJ\mathcal{O}}\neq0$. However, the fact
that the OPE combination must vanish is a necessary but not sufficient
condition to show that these points are at infinite distance. To complete the
proof, we require information about their neighborhood, and what happens when
the higher-spin symmetry is weakly broken.

\subsection{Weakly-broken Higher-Spin Symmetries} \label{sec:weakly-broken}

In this section we move slightly away from the higher-spin point. This is, we
assume that the higher-spin symmetry is weakly broken by a small parameter $g$
(related, but not to be confused with the marginal coupling $t$ used in conformal
perturbation theory). This is, for the divergence of the higher-spin operators
we have:
\begin{equation} \label{eq:HS-breaking}
	\partial \cdot J_{\ell} = g \, K_{\ell-1} \, .
\end{equation}
Here, we have introduced an operator $K_{\ell-1}$ controlling the higher-spin
symmetry breaking. This operator is a primary of twist $d$ and spin $\ell-1$
only at the HS point, i.e., at leading order in $g\to 0$ \cite{Anselmi:1998ms,
Henn:2005mw, Maldacena:2012sf}. In fact, the two multiplets recombine into a
larger multiplet with primary $J_{\ell}$---which is no longer conserved due to
the anomalous dimension---and $K_{\ell-1}$ is a mere descendant of the would-be
current away from the higher-spin point \cite{Rychkov:2015naa}. We take this
operator to be in an orthonormal basis, also at leading order in this limit.
This in turn fixes the definition of the higher-spin symmetry breaking
parameter $g$ via equation \eqref{eq:HS-breaking}.

Breaking the HS symmetry, the decoupled sector is not free anymore, and
$K_{\ell-1}$ can be thought of as specifying the type of interactions that are
turned on. However, we will not assume that the CFT data enjoys an expansion in
powers of $g$ around the HS point. The reason is that $g$ is not necessarily
the coupling constant of those interactions. For instance, it could \emph{a
priori} happen that $\partial \cdot J_{\ell}$ vanishes at tree level, and the
breaking occurs only at higher-loop orders. In this case, the CFT data would
generically have an expansion in fractional powers of $g$. Here, we will only
assume that the CFT data remains finite as the HS point is approached. This is
closely related to the normalization of operators such as $J_\ell$,
$K_{\ell-1}$, and $\mathcal{O}$. Normalizations in which we allow the two-point
functions to diverge as $g\to 0$ can lead to similar, unphysical, divergences
in the three-point functions. Nevertheless, this cannot happen when operators
are properly normalized. In this case the three-point-function coefficients are
physical and should remain finite in any \emph{bona fide} CFT like those
associated with HS points of the conformal manifold.

As pioneered in reference \cite{Anselmi:1998ms}, let us now take the double
divergence of the two-point function of two higher-spin operators, and use the
broken Ward identity defined in equation \eqref{eq:HS-breaking}:
\begin{equation}
	\partial_1 \partial_2 \cdot \left\langle J_{\ell}(x_{1}) J_{\ell}(x_{2})\right\rangle = g^{2} \left\langle K_{\ell-1}(x_{1}) K_{\ell-1}(x_{2}) \right\rangle \, .
\end{equation}
The LHS of this expression can be directly computed using the form of the
two-point function as an expansion in terms of the conformal structures, see
equation \eqref{2pt-function}. For small anomalous dimensions, $\gamma_\ell =
\tau_{\ell}-(d-2)\ll1$, it is at leading order proportional to a
spin-$(\ell-1)$ two-point function \cite{Skvortsov:2015pea, Giombi:2016hkj}.
Since $K_{\ell-1}$ is a conformal primary for $g\to0$, the RHS is also
proportional to the two-point function of a spin-$(\ell-1)$ conformal primary
in that limit. We then conclude that, at leading order, the anomalous dimension
is given by:
\begin{equation} \label{eq:HS-anomalous-dimension}
	\gamma_{\ell} \sim g^{2}\,,\qquad \text{as } g\to 0 \, ,
\end{equation}
where we are ignoring numerical prefactors whose precise values can be found in
references \cite{Skvortsov:2015pea, Giombi:2016hkj}. In this way we see that
the first contribution to the anomalous dimension of the higher-spin operators
always appears at order $g^{2}$ in the higher-spin symmetry-breaking parameter.
This result is sometimes referred to as \emph{Anselmi's trick}, who first used
the broken Ward identity to simplify the computations of the anomalous
dimensions of the Konishi current and other spinning operators for
four-dimensional $\mathcal{N}=4$ super-Yang--Mills theory\cite{Anselmi:1998ms}.

We can apply similar arguments to the three-point function $\langle J_{\ell}
J_{\ell} \mathcal{O} \rangle$. Let us first take the divergence with respect to
the second higher-spin operator and use equation \eqref{eq:HS-breaking}:
\begin{equation} \label{eq:single-divergence}
	\partial_2 \cdot \left\langle J_{\ell}(x_{1}) J_{\ell}(x_{2}) \mathcal{O}(x_3) \right\rangle = 
	g \left\langle J_{\ell}(x_{1}) K_{\ell-1}(x_{2}) \mathcal{O}(x_3) \right\rangle \, .
\end{equation}
As for the two-point function, at leading order the divergence of the currents
matches with the correlator of spinning primaries. Something similar happens
for the three-point function in the RHS, and we see that the leading-order
corrections to the current-conservation condition, which led to
\eqref{eq:recursion-relation}, are controlled by the behavior of $\left\langle
J_{\ell}K_{\ell-1}\mathcal{O}\right\rangle$ close to the HS point.

Similarly, a weakly-broken version of the integrated Ward identity goes as
follows: because of equation \eqref{eq:HS-breaking}, the charges defined in
equation \eqref{charges} are no longer conserved, and as a result their
correlators depend on the codimension-one surface $\Sigma$. One can however
define the action of quasi-conserved charges on a given operator as the result
of the integral when the surface $\Sigma$ enclosing its position shrinks to
zero size \cite{Maldacena:2012sf}. This action is constrained by twist
conservation as it was the case when the symmetry was exactly preserved. By
virtue of Stokes' theorem, one then gets integrated Ward identities similar to
the conserved case, but with an extra contribution from the correlator where
$J_\ell$ has been replaced by $\partial \cdot J_\ell = g K_{\ell-1}$,
integrated over the volume enclosed by $\Sigma$. As happens for the
current-conservation condition, the leading-order correction to the integrated
Ward identity for $\left<J_\ell J_\ell \mathcal{O}\right>$ is again controlled
by the behavior of $\left\langle J_{\ell}K_{\ell-1}\mathcal{O}\right\rangle$
close to the HS point.

Putting these two results together and taking into account equation \eqref{JJO
vanishes}, we get that for a weakly-broken HS symmetry we have 
\begin{equation} \label{JJO correction}
    C_{JJ\mathcal{O}} \sim g \, C_{JK\mathcal{O}} \qquad \text{ as } g\to 0 \, ,
\end{equation}
where $C_{JK\mathcal{O}}$ is a certain linear combination of the OPE
coefficients appearing in $\left<J_\ell K_{\ell-1} \mathcal{O}\right>$ as
defined in equation \eqref{JKO}. 

If $C_{JK\mathcal{O}}$ is non-vanishing at the HS point, we see that the OPE
combination behaves as $C_{JJ\mathcal{O}} \sim g$ when $g\to 0$. Using equation
\eqref{eq:HS-anomalous-dimension}, this would mean that at leading order, we
would have $C_{JJ\mathcal{O}}\sim \gamma^{1/2}$. Following the CFT Distance
Criterion discussed in Section \ref{sec:finite-vs-infinite}, this would result
in HS points being at finite distance with respect to the trajectory defined by
the marginal operator $\mathcal{O}$. For the first part of the CFT Distance
conjecture to be true, the OPE combination $C_{JK\mathcal{O}}$ must always be
trivial at the HS point. We now show this is indeed the case.

\subsubsection{The Correlator $\langle J_\ell K_{\ell-1}\mathcal{O}\rangle$ at the HS Point}

To show that at the higher-spin point the correlator $\langle J_\ell
K_{\ell-1}\mathcal{O}\rangle$ vanishes, we will go through a similar procedure
to the one we used for the three-point function $\langle J_\ell
J_{\ell}\mathcal{O}\rangle$, and begin with the current-conservation condition.
We now have to take the divergence of a type-$(\ell,\ell-1,0)$ correlator with
respect to the first operator, from which we obtain a three-point function of
type-$(\ell-1,\ell-1,0)$. Using that $K_{\ell-1}$ is a conformal primary at
this point, and as explained in Section \ref{sec:spinning-correlators}, both
correlators involve $\ell$ independent conformal structures. As before, this
enable us to find an homogeneous system of $\ell$ linear constraints for $\ell$
variables $C_{JK\mathcal{O}}^{n}$. 

If all the equations were independent, one would automatically get that
$C_{JK\mathcal{O}}^{n}=0$. Unfortunately, we have checked that the system is
not closed for $2 \leq \ell \leq 10$ and for any $d$. In general, we therefore
expect that this happens for any spin and any number of spacetime dimensions.
Despite this, we show in the Appendix \ref{sec:appendix-A} that the constraints
can always be used to fix most of the coefficients, and we get a similar result
as for $C_{JJ\mathcal{O}}^n$:
\begin{equation} \label{eq:recursion-relation-2}
		C_{JK\mathcal{O}}^{n} = \tilde v^{n} \, C_{JK\mathcal{O}}^{\ell-1} \, ,\qquad n=0,\dots,\ell-1\,,
\end{equation}
where the components of the vector, denoted by $\tilde{v}^n\in\mathbb{R}$, are normalized such that
$\tilde{v}^{\ell-1}=1$. In appendix \ref{sec:appendix-A}, we argue that,
similarly to what we have done for the case of $\left< J_\ell J_\ell
\mathcal{O}\right>$ above, the current-conservation constraint imposes a
recursion relation in the space of OPE coefficients $C_{JK\mathcal{O}}^n$. We
have not found a closed-form expression for the coefficients $\tilde{v}^n$, but
they can easily be solved for a given spin programmatically. For instance, for
$\ell=4$ it is straightforward to check that the solution is given by:
\begin{equation}
	\tilde{v}^n = \left( \frac{-24}{(d-2)d(d+2)} \, ,\, \frac{36}{(d-2)d} \, ,\, \frac{-12}{d-2} \, ,\, 1 \right) \, .
\end{equation}
Once again, the particular values of $\tilde{v}^n$ is not important, only that
that the current-conservation condition fixes all the coefficients but one.

\medskip

To show that this remaining coefficient is in fact trivial, we again need to
impose the integrated Ward identity. Starting with the analogue of equation
\eqref{ward-contact-terms}, we integrate the current over a manifold enclosing
the point $x_2$ with boundary $\Sigma$, and we obtain:
\begin{equation} \label{eq:Ward}
	\int_{\Sigma} \left\langle \check{J}_{\ell}(x_1,\xi_1) K_{\ell-1}(x_2,\xi_2) \mathcal{O}(x_3) \right\rangle \propto \left\langle \left[Q_{\ell}(\xi_1) ,K_{\ell-1}(x_2,\xi_2)\right] \mathcal{O}(x_3) \right\rangle \, ,
\end{equation}
and $\check{J}_\ell$ is the current with the first index in the direction
normal to the surface. As we did for the three-point function $\left\langle
J_{\ell} J_{\ell} \mathcal{O} \right\rangle$, taking $\Sigma$ to pick the
$\xi_1$-direction, the LHS can be expanded in terms of the conformal
structures. Taking into account equation \eqref{eq:recursion-relation-2}, we
then have:
\begin{equation}
		\int_{\Sigma} \left\langle J_{\ell}(x_1,\xi_1) K_{\ell-1}(x_2,\xi_2) \mathcal{O}(x_3) \right\rangle = C_{JK\mathcal{O}}^{\ell-1} \int_{\Sigma} \left( \sum_{n=0}^{\ell} \tilde v^{n} \widetilde{\Theta}_{n}(x_i;\xi_1,\xi_2) \right) \, .
\end{equation}
As discussed in Appendix \ref{app:surface-integrals}, the same choice of
polarization vector and surface we used for $\left\langle J_{\ell} J_{\ell}
\mathcal{O} \right\rangle$ again lead to a non-vanishing integral. However, a
crucial difference is that the commutator in equation \eqref{eq:Ward} is not
trivial in this case. Indeed, due to twist conservation, only twist $\tau=d$
operators can appear in the commutator from the relation given in equation
\eqref{Qkcomm}. Since marginal operators also have twist $d$, it can in
principle be contained in the commutator, and we cannot conclude that the
three-point function $\left\langle J_\ell K_{\ell-1}\mathcal{O}\right\rangle$
vanishes as in the previous subsection.

\medskip

To do so, we need to invoke an ingredient that we have not used so far, namely
the presence of the energy-momentum tensor. Indeed, assuming a local CFT is
what makes the CFT distance conjecture part of the holographic Swampland
Program, as it implies the presence of dynamical gravity in the bulk. Note that
here we do not assume a weakly-coupled gravity dual, and will only use the
interplay of the energy-momentum tensor with the higher-spin algebra. As
alluded to in Section \ref{sec:charge-conservation}, it can be used to
constrain the operators appearing in the commutators involving higher-spin
charges \cite{Maldacena:2011jn}.

In particular, our goal is to show that $\left<J_\ell K_{\ell-1}
\mathcal{O}\right>\neq 0$ is inconsistent with a charge-conservation identity.
We will achieve this by considering the action of $Q_{\ell}$ on the three-point
function $\left\langle T_2 K_{\ell-1} \mathcal{O} \right\rangle$:
\begin{equation} \label{charge-condition}
   \left< \left[ Q_{\ell} , t_2 k_{\ell-1} \mathcal{O} \right] \right> = \left< \left[ Q_{\ell} , t_2 \right] k_{\ell-1} \mathcal{O} \right> + \left< t_2 \left[ Q_{\ell} , k_{\ell-1} \right] \mathcal{O} \right> + \left< t_2 k_{\ell-1} \left[ Q_{\ell} , \mathcal{O} \right] \right> = 0 \, ,
\end{equation}
which can be understood as integrating equation \eqref{ward-contact-terms} over
all of spacetime. We recall that we use the notation---explained in Section
\ref{sec:charge-conservation}---where lowercase operators indicate that the
indices are all taken in the $x^-$-direction, e.g. $t_2=T_{--}$ for the
energy-momentum tensor, and the HS charges are defined as in equation
\eqref{charges-minus}. 

Let us first discuss the action of one of these charges on the energy-momentum
tensor:
\begin{equation}
    \left[Q_\ell , t_2\right] = \sum_{s=0}^{\ell} \alpha_s \, \partial^{\ell +1 -s} j_s \, ,
\end{equation}
Recall that $\partial$ denotes the partial derivative with respect to the
$x^-$-direction. Even though it is allowed by twist and spin conservation,
notice that we have not included $j_{\ell+1}$. The reason is that, were it to
appear in the commutator then $Q_\ell$ would not be translation invariant
\cite{Maldacena:2011jn}. Note that we are not assuming uniqueness of the 
spin two conserved current, in general $j_2$ could contain contributions from 
both the energy-momentum tensor $t_2$, and any other spin-two current.

The action of $Q_{\ell}$ on $k_{\ell-1}$ and $\mathcal{O}$ leads to similar
expressions, see equation \eqref{Qkcomm}:
\begin{equation}\label{[Q,O]-[Q,K]}
\begin{split}
    \left[Q_\ell , k_{\ell-1}\right] &= \sum_{s=0}^{2\ell -2} \eta_s \, \partial^{2\ell -2 -s} k_{s}^{\prime} \, , \\
    \left[Q_\ell , \mathcal{O} \right] &= \sum_{s=0}^{\ell-1} \beta_s \, \partial^{\ell -1-s} k_{s}^{\prime} \, .
\end{split}
\end{equation}
The primed operators $k_{s}^\prime$ are there to remind us that they might be
different from unprimed operators $k_{\ell-1}$ we have considered so far.
Furthermore, the twist-$d$ scalar operators $k_0^\prime$ might be different
from the marginal operator $\mathcal{O}$ we started with. 

Plugging these commutators into equation \eqref{charge-condition}, we then
obtain:
\begin{equation} \label{charge-conservation-general}
   \sum_{s=0}^{\ell} \alpha_s \, \partial^{\ell +1 -s}_{1} \left< j_s k_{\ell-1} \mathcal{O} \right> + \sum_{s=0}^{2\ell -2} \eta_s \, \partial^{2\ell -2 -s}_{2} \left< t_2 k^{\prime}_{s} \mathcal{O} \right> + \sum_{s=0}^{\ell-1} \beta_s \, \partial^{\ell -1-s}_{3} \left< t_2 k_{\ell-1} k_{s}^{\prime} \right> = 0 \, ,
\end{equation}
where $\partial_i$ denotes the partial $x^-$-derivative with respect to the
position of the $i$-th operator. Notice that the first sum contains a term with 
$\alpha_\ell \neq 0$ and $\left<j_\ell k_{\ell-1} \mathcal{O}\right>$, if the charge-conservation 
identity imposes that this term must vanish, then we obtain the desired result.

Performing this analysis for general spin $\ell$ is quite involved, but it
becomes more manageable for $\ell=4$. In this case we have:
\begin{equation} 
   \sum_{s=0}^{4} \alpha_s \, \partial^{5 -s}_{1} \left< j_s k_{3} \mathcal{O} \right> + \sum_{s=0}^{6} \eta_s \, \partial^{6-s}_{2} \left< t_2 k^{\prime}_{s} \mathcal{O} \right> + \sum_{s=0}^{3} \beta_s \, \partial^{3-s}_{3} \left< t_2 k_{3} k_{s}^{\prime} \right> = 0 \, .
\end{equation}
We then have to analyze correlators of type $\left< J_s K_{3} \mathcal{O} \right>$, $\left< T_2 K_{s} \mathcal{O} \right>$ and $\left< T_2 K_{3} K_{s}\right>$. Using the current-conservation condition and integrated Ward identities, we conclude that:
\begin{equation}
\begin{split}
    \left< J_s K_{3} \mathcal{O} \right> &= 0 \qquad \text{for } s=2,3 \, ; \\
    \left< T_2 K_{s} \mathcal{O} \right> &= 0 \qquad \text{for } s=1, \dots, 6 \, ; \\
    \left< T_2 K_{3} K_{s} \right> &= 0 \qquad \text{for } s=0 \, .
\end{split}
\end{equation}
This can be shown as we did for $\left<J_\ell J_\ell \mathcal{O}\right>$: the current conservation condition reduces the number of independent OPE coefficients to one, and the integrated Ward identity then imposes that the remaining OPE coefficient must vanish. For the case $\left< T_2 K_{1} \mathcal{O} \right>$ it is crucial to use that $T_2$ is the energy-momentum tensor. Only in this case $Q_2$ is the charge generating translations in the minus direction, \textbf{$Q_2=P_{-}$}, which guarantees that even though allowed by twist conservation, $K_1$ does not appear in $[Q_2,\mathcal{O}]$. The results for the current-conservation conditions and the integrals for the integrated Ward identities can be found in appendices \ref{sec:appendix-A} and \ref{app:surface-integrals}.

Using this, the charge-conservation condition reduces to
\begin{equation} \label{charge-conservation-final}
\begin{split}
    0 = \, &\alpha_0 \, \partial^{5}_{1} \left< j_s k_{0} \mathcal{O} \right> + \alpha_1 \, \partial^{4}_{1} \left< j_1 k_{3} \mathcal{O} \right> + \alpha_4 \, \partial_{1} \left< j_4 k_{3} \mathcal{O} \right> + \, \eta_0 \, \partial^{6}_{2} \left< t_2 k^{\prime}_{0} \mathcal{O} \right> \\
     & + \beta_1 \, \partial^{2}_{3} \left< t_2 k_{3} k_{1}^{\prime} \right> + \beta_2 \, \partial_{3} \left< t_2 k_{3} k_{2}^{\prime} \right> + \beta_3 \, \left< t_2 k_{3} k_{3}^{\prime} \right>\, .
\end{split}
\end{equation}
The last three terms are more complicated to analyze, since the current-conservation condition allows for more than one conformal structure. However, including them in the charge-conservation condition becomes more manageable when coordinates are taken in the light-cone plane, $x_i=(x_{i}^{-},x_{i}^{+},\mathbf{0})$, since only one linear combination of the conformal structures of these correlators contributes in this case.\footnote{It is not guaranteed that this linear combination survives the current-conservation condition. However, that would not change our results, but make them easier to derive.} Moreover, the spacetime dependence of all the conformal structures simplify drastically. Imposing this charge-conservation condition for any $x_{i}^{-}$ and $x_{i}^{+}$, we find that there is no solution if $\left< j_4 k_{3} \mathcal{O} \right>\neq0$, thus arriving to the desired result:
\begin{equation}
    \left< J_4 K_{3} \mathcal{O} \right> = 0 \, .
\end{equation}

\medskip

This concludes the analysis of the correlator
$\left<J_{\ell}K_{\ell-1}\mathcal{O}\right>$ at the HS point. At least for the
case $\ell=4$, we have been able to show that this correlator must vanish for any marginal operator $\mathcal O$. This
in turn implies that
\begin{equation} \label{JKO vanishes}
		C_{JK\mathcal{O}}^{\text{HS}} = 0\,,
\end{equation}
as we wanted. Following the same procedure as for $\ell=4$,
we expect the same result to hold for $\ell>4$. In fact, this has to be the case for any other HS conserved current coexisting with
$J_4$ to generate the higher-spin algebra, otherwise the conformal manifold
would stop making sense as a metric space. We come back to this point in the
next section.

\subsection{Infinite Distance and Exponentially Conserved Currents} \label{sec:exponential}

In this section, we have so far shown that by using various properties of the
higher-spin algebra, two of the three-point functions involving marginal
operators and a HS current must always be trivial at the HS point:
\begin{equation}\label{CJJO=0=CJKO}
		\text{HS point:}\qquad \langle J_{\ell}J_\ell\mathcal{O}\rangle=0=\langle J_{\ell}K_{\ell-1}\mathcal{O}\rangle\,.
\end{equation}
Following the CFT Distance Criterion defined in Section
\ref{sec:finite-vs-infinite}, the first term is
a necessary condition for HS points to be at infinite distance in the conformal
manifold. On the other hand, from the discussion around equation
\eqref{eq:single-divergence}, the fact that second vanishes forbids
contributions to the OPE combination $C_{JJ\mathcal{O}}$ at linear order in the
parameter $g$ for a weakly-broken HS symmetry.

By the same token, one can reach a similar conclusion for
$C_{JK\mathcal{O}}(g)$ away from the HS point. Taking the divergence with
respect to $J_\ell$ and using equation \eqref{eq:HS-breaking} we have
\begin{equation} \label{weakly-broken-JKO}
    \partial_1 \cdot \left<J_{\ell}(x_1)K_{\ell-1}(x_2)\mathcal{O}(x_3)\right> = g \left<K_{\ell-1}(x_1)K_{\ell-1}(x_2)\mathcal{O}(x_3)\right> \, .
\end{equation}
The leading order correction to the current-conservation condition for
$\left<J_{\ell}K_{\ell-1}\mathcal{O}\right>$ is controlled by
$\left<K_{\ell-1}K_{\ell-1}\mathcal{O}\right>$ at the HS point. The same
observation can be reached for the leading correction to the integrated Ward
identity and the charge-conservation condition \cite{Maldacena:2012sf}.
Similarly to what we obtained in the previous section, and taking into account
equation \eqref{JKO vanishes}, we now have
\begin{equation} \label{JKO correction}
    C_{JK\mathcal{O}} \sim g \, C_{KK\mathcal{O}} \qquad \text{ as } g\to 0 \, ,
\end{equation}
where $C_{KK\mathcal{O}}$ is again a certain linear combination of the OPE
coefficients appearing in $\left<K_{\ell-1}K_{\ell-1}\mathcal{O}\right>$.

Since both $K_{\ell}$ and $\mathcal{O}$ are assumed to be in an orthonormal
basis, the correlator $\left<K_{\ell-1}K_{\ell-1}\mathcal{O}\right>$ cannot
blow up as $g\to 0$---see the discussion below equation \eqref{eq:HS-breaking}.
In general, we therefore have
\begin{equation}
    \left<K_{\ell-1}K_{\ell-1}\mathcal{O}\right> \sim g^{2a} \qquad \text{ with } a\geq 0 \, \text{ as } g\to 0 \, .
\end{equation}
By this equation, we mean that the $\ell$ OPE coefficients of this correlator,
$C_{KKO}^{n}$, have this behavior as $g\to 0$. Combining this with equations
\eqref{JJO correction} and \eqref{JKO correction}, we have
\begin{equation} \label{JJO behavior}
	C_{JJ\mathcal{O}} \sim g^{2(1+a)} \qquad \text{ with } a\geq 0 \, \text{ as } g\to 0 \, 
\end{equation}
for the OPE combination controlling the conformal-perturbation-theory equation
\eqref{eq:master-eq}.\footnote{Strictly speaking, it could happen that the
coefficients $C_{JJ\mathcal{O}}^{\,n}$ conspire in such a way that
leading correction vanishes for the linear combination controlling
conformal perturbation theory, $C_{JJ\mathcal{O}}$. Equation \eqref{JJO
behavior} would then depend on another exponent $a^{\prime}>a$, but the
conclusion would be unchanged.} Given the behavior of the anomalous dimension
in equation \eqref{eq:HS-anomalous-dimension}, we finally reach the conclusion
\begin{equation}\label{final-result}
  C_{JJ\mathcal{O}} \sim \gamma_{\ell}^{1+a} \qquad \text{ with } a\geq 0 \, \text{ as } \gamma_\ell \to 0 \, .
\end{equation}
Since we have made no assumption on the marginal operator $\mathcal{O}$, this
behavior is independent of the deformation away from the higher-spin point, and
is precisely the CFT Distance Criterion for an infinite-distance point in the
conformal manifold, as explained in Section \ref{sec:finite-vs-infinite}.
Furthermore, we have not made any assumptions on the operators $J_\ell$ either,
apart from the fact that they become conserved at the HS point. This then shows
that for local CFTs in any dimension, the first part of the CFT Distance
Conjecture is correct:
\begin{center}
		\emph{all higher-spin points are at infinite distance in the conformal manifold}.
\end{center}

Strictly speaking, in Section \ref{sec:weakly-broken} we have only proven this
for spin-four currents. However, as discussed below the CFT Distance Criterion
in Section \ref{sec:finite-vs-infinite}, it is enough to show that a single
higher-spin current places the HS point at infinite distance in the conformal
manifold. The rest of the tower should behave accordingly, and we must also
have $\left<J_{\ell} K_{\ell-1} \mathcal{O}\right>=0$ at the HS point for any
other current. Furthermore, since there is always a spin-four conserved current
generating the higher-spin algebra \cite{Maldacena:2011jn,
Stanev:2013qra, Boulanger:2013zza, Alba:2013yda, Alba:2015upa}, the case $\ell=4$
suffices to establish that any HS point is at infinite distance.

\medskip

It is also interesting to elaborate more on the condition for the higher-spin
currents to become conserved exponentially with the distance. This would be the
case if $a=0$ in equation \eqref{final-result}. In other words, having an
exponentially conserved HS current with the distance as the HS point is
approached is linked to having:
\begin{equation} \label{exp-behavior}
    \left<K_{\ell-1}K_{\ell-1}\mathcal{O}\right>_{\text{HS}} \neq 0 \, .
\end{equation}
In particular, this does not mean that all the OPE coefficients
$C_{KK\mathcal{O}}^n$ are non-vanishing as the HS point. We only need to
require this for the OPE combination $C_{KK\mathcal{O}}$ controlling the order
$g^2$ correction to $C_{JJ\mathcal{O}}$. It is intriguing that the condition
for the exponential behavior precisely involves the last correction to
$C_{JJ\mathcal{O}}$ that can be addressed using Anselmi's trick, i.e. the
weakly-broken Ward identity \eqref{eq:HS-breaking}.

Having $\left<K_{\ell-1}K_{\ell-1}\mathcal{O}\right> \neq 0$ at the infinite
distance HS point might seem a bit puzzling. Indeed, one of the corollaries of
the CFT Distance Criterion is that having an operator with non-vanishing three
point function with $\mathcal{O}$ is generically enough to imply that the point
is at finite distance. However, let us recall that $K_{\ell-1}$ is not a
conformal primary outside the HS point, while the CFT Distance Criterion
assumes this for the operator under consideration, and equation
\eqref{eq:master-eq} therefore does not apply to the descendant $K_{\ell-1}$.
The analogous equation, obtainable through conformal perturbation theory
applied to this non-primary operator, should be such that the behavior of
$C_{KK\mathcal{O}}$ close to the HS point must place it at infinite distance.
In fact, in our case this is guaranteed by the fact that, when going away from
the HS point, the current and $K_{\ell-1}$ recombine into a larger multiplet,
and their behavior as we approach the free point must be related.

If equation \eqref{exp-behavior} is satisfied, then the exponential decay rate
$\alpha_{\chi}^{(\ell)}$ appearing in equation \eqref{exponential-SDC} is
controlled by the OPE combination $C_{KKO}$ at the HS point. Schematically, we
have:
\begin{equation}
    \alpha_{\chi}^{(\ell)} \sim \left<K_{\ell-1}K_{\ell-1}\mathcal{O}\right>_\text{HS} \, ,
\end{equation}
where the numerical coefficients and the OPE combination, $C_{KK\mathcal{O}}$,
can be fixed by the weakly-broken HS symmetry constraints. In fact, this result
can be brought into a more suggestive form. Taking an orthonormal basis for the
set of marginal operators $\mathcal{O}_i$ and generalizing the evolution
equation \eqref{eq:master-eq} we have
\begin{equation} \label{scalar-charge-to-mass}
    \frac{\partial_i \gamma_\ell}{\gamma_\ell} \sim \left<K_{\ell-1}K_{\ell-1}\mathcal{O}_{i}\right>_\text{HS} \, .
\end{equation}
An orthonormal basis for the marginal operators corresponds to an orthonormal
basis for the tangent bundle of the conformal manifold. Therefore, equation
\eqref{scalar-charge-to-mass} gives the components of what is defined as the
scalar charge-to-mass ratio vector used for the convex-hull formulation of the
Swampland Distance Conjecture \cite{Calderon-Infante:2020dhm}. This equation is
then telling us that, close to the HS point, this information is encoded in
this particular three-point function.

\medskip

Showing the condition given in equation \eqref{exp-behavior} for a single spin,
e.g. $\ell=4$, is not enough to show an exponential behavior for all off them.
There is \emph{a priori} nothing wrong with having an infinite-distance point
for which the different HS currents become conserved at different rates along a
given trajectory in the conformal manifold. The only thing that is required by
consistency of the conformal manifold as a metric space is that
$\gamma_\ell(t)\to 0$ as $t\to\infty$ for all of them.

Despite this, there might be a way of translating the result for a single spin
to the rest of the tower. The idea is to exploit a relation that we have not
discussed yet. Even though the HS symmetry-breaking parameter $g$ is not the
same as the distance $t$ in the conformal manifold, they are related. Indeed,
having shown the exponential behavior for the anomalous dimension of a given
spin $\ell$, and using equation \eqref{eq:HS-anomalous-dimension} we get the
relation:
\begin{equation}
    g_{\ell}^2 \sim e^{- \alpha_{\chi}^{(\ell)} \, t} \qquad \text{ as } t\to \infty \, .
\end{equation}
Here we have added a subscript $\ell$ to the HS symmetry breaking parameter to
make clear that \emph{a priori} one could expect the parameters to be different
for the breaking of the conservation condition of different HS operators.
However, coming back to the underlying picture of introducing an interaction
with a small coupling, we indeed expect them to be related in a power-like
fashion. This is, we expect each of these conservation conditions to be broken
at a given loop order. Note however that this could be avoided if the
conservation condition for some HS currents is broken only by non-perturbative
effects.

Following this reasoning, the exponential behavior for a given spin would carry
over to the others, even though the exponential decay rate might still depend
on $\ell$. It would be interesting to show that, as it happens for super-Yang--Mills
theories, all the HS conservation conditions are broken at the same order, i.e,
$g_\ell=g$. In fact, it seems plausible that this is imposed by consistency of
the HS symmetry breaking, since this symmetry relates the different HS
currents. If this is case, one would not only conclude that it is enough to
show the exponential behavior for a single spin, but also that the exponential
rate $\alpha_\chi$ does not depend on $\ell$ and one has a full tower of HS
currents becoming exponentially conserved at the same rate.

\section{Flavor Enhancements at (In)Finite Distances}\label{sec:flavor-infinite-distance}

In the previous section, we have used properties of the higher-spin algebra to
show that HS points are at infinite distance. Although we focused on
higher-spin currents, many aspect of this machinery turns out to also be valid
for $\ell=1$. In the conformal manifold, this corresponds to regimes where a
new conserved flavor current appears in the spectrum. These special points need
not automatically be associated with higher-spin points, nor with infinite
distances. A classical family of examples is $(\mathcal{N}=1)$-preserving
deformations of four-dimensional $\mathcal{N}=4$ super-Yang--Mills
\cite{Leigh:1995ep}, where new R-symmetry currents are accompanied with extra
supercharges, see e.g. reference \cite{Cordova:2016emh} for an analysis of the
possible multiplet recombination rules. These enhancements are associated with
superpotential deformations, and are at finite distance
\cite{Perlmutter:2020buo}.

On the other hand, not all flavor-enhancement point are at finite distance. In
$\mathcal{N}=2$ four-dimensional theories, representation-theoretic arguments
show that spin-one would-be-conserved current are in the same long multiplet as
would-be higher-spin currents. At threshold, the long multiplet then splits
into (at least) two short multiplets and both the flavor and HS currents become conserved
\cite{Dolan:2002zh, Beem:2014zpa, Cordova:2016emh}. The flavor-enhancement
points are therefore also higher-spin points, and always at infinite-distance
in the conformal manifold from the results of the previous section. This is
related to the decomposability conjecture \cite{Beem:2014zpa}, stating that
non-trivial $\mathcal{N}=2$ conformal manifolds can only be obtained by
performing a conformal gauging of the flavor symmetry of isolated theories.
The decoupled sector is then made out of the spectrum of free vector
multiplets.

One can then ask whether a similar fate awaits more general cases, where we
only assume that there exist a neighborhood of the flavor-enhancement point
such that the broken Ward identity is valid:
\begin{equation}
		\partial_\mu J^\mu = g\,K_0 \, .
\end{equation}
At the point where $g=0$, $K_0$ is a scalar conformal primary with
$\Delta=d$, but develops an anomalous dimension elsewhere. The two operators $J_1$
and $K_0$ recombine into a larger multiplet away from that point, where $K_0$
is a conformal descendant \cite{Rychkov:2015naa}.

The first step of our argument required us to show that, when $g=0$, the OPE
combination vanishes: $C_{JJ\mathcal{O}}=0$. The current-conservation condition
applied to the correlator $\langle J_1J_1\mathcal{O}\rangle$ again fixes the
two OPE coefficents to be proportional to one another:
\begin{equation}\label{vn-spin-1}
	C_{JJ\mathcal{O}}^0 = \frac{1}{d} C_{JJ\mathcal{O}}^1\,.
\end{equation}
But, contrary to cases where $\ell>1$, we cannot use the integrated Ward
identity to conclude that the remaining coefficient must vanish. Indeed, as
shown in Appendix \ref{sec:appendix-A} for the choice of surface $\Sigma$
discussed in Section \ref{HS-conserved-currents-condition}, the integral
vanishes and we cannot conclude that $C_{JJ\mathcal{O}}=0$ in this case.

It is nevertheless tempting to assume that the OPE combination should vanish at
the flavor-enhancement point. Were it not the case, taking $\gamma^0_{\ell=1}=0$
as the reference point in equation \eqref{eq:solution}, the anomalous dimension 
would become negative, $\gamma \simeq - C_{JJ\mathcal{O}}
\, t<0$, and violate unitarity bounds. Since it is at least in principle
possible to deform away from the flavor-enhancement point in the directions
associated with both $\pm\mathcal{O}$, corresponding to the two directions away
from the point along the line parameterized by $t$, one of them always lead to
a violation of unitarity. A possible escape could be that somehow conformal
perturbation theory breaks down in that direction, leaving us with only a
semi-infinite critical line.

For now, let us thus assume that the OPE combination indeed vanishes:
$C_{JJ\mathcal{O}}=0$. We are then led to consider the correlator $\langle
J_1K_0\mathcal{O}\rangle$ at the point $g=0$. The current-conservation
condition is automatically satisfied, and using the integrated Ward identity we
obtain:
\begin{equation}\label{integrated-ward-spin-1}
		\int_\Sigma dx^\mu\, \langle J_{1,\mu}(x_1) K_{0}(x_2)\mathcal{O}(x_3) \rangle \propto \langle [Q_1,K_0(x_2)]\mathcal{O}(x_3)\rangle \, .
\end{equation}
The integral on the LHS can be expanded in terms of the single type-$(1,0,0)$
conformal structure, and, as shown in Appendix \ref{app:surface-integrals}, the
result is non-vanishing in any dimension. 

Whether the OPE combination $C_{JK\mathcal{O}}$ is trivial---and by extension
whether the flavor-enhancement point is at infinite distance via the CFT
Distance Criterion---is therefore decided by the commutator $[Q_1,K_0]$. If any
of the marginal operators belong to the same multiplet as the twist-$d$
operator, $[Q_1,K_0]\sim\mathcal{O}$, then the two-point function on the RHS of
equation \eqref{integrated-ward-spin-1} is non-vanishing and
$C_{JK\mathcal{O}}\neq0$. The associated limiting point is then at finite
distance. If on the other hand none of the marginal operators share a multiplet
with $K_0$, the RHS of equation \eqref{integrated-ward-spin-1} is trivial and
the flavor-enhancement point is at infinite distance---provided that
$C_{JJ\mathcal{O}}=0$ at that point, as discussed above.

Let us note that the charge-conservation identity applied in Section
\ref{sec:weakly-broken} for the case of HS symmetry does not help in this case.
Precisely for $\ell=1$, $j_1$ cannot appear in $[Q_1,j_2]$, since $Q_1$
has vanishing conformal dimension. This is the usual statement that the
energy-momentum tensor should be invariant under any global symmetry. We thus
have to set $\alpha_1=0$ in equation \eqref{charge-conservation-general} and the
correlator of interest $\left<j_1 K_0 \mathcal{O}\right>$ does not appear in the charge-conservation condition.

In addition, the unitarity argument that we presented in favor of
$C_{JJ\mathcal{O}}=0$ at the flavor-enhancement point does not work either. If
we have $C_{JK\mathcal{O}}\neq 0$, then the behavior of the OPE combination is
given by $C_{JJ\mathcal{O}}\sim g \sim \sqrt{\gamma_1}$, which leads to
$\gamma_1 \sim t^2$ upon using equation \eqref{eq:master-eq}. This is at finite
distance, but the anomalous dimension is not linear with the Zamolodchikov
distance $t$, as it is the case for $C_{JJ\mathcal{O}}\neq 0$ at the
flavor-enhancement point. Therefore, this would not lead to a violation of the
unitarity bound in any of the two directions departing from the
flavor-enhancement point along the line parametrized by $t$. 

\medskip

The condition we have found for the flavor-enhancement point to be at finite
distance is of course very intuitive. It is equivalent to the statement that
the marginal operators are charged under the symmetry. We could have
alternatively chosen another integration domain so that equation
\eqref{integrated-ward-spin-1} is written in terms of the charges acting on
$\mathcal{O}$, which would have led us to the condition $[Q_1,\mathcal{O}]\sim
K_0$. For a deformation---marginal or not---to preserve a symmetry the
associated operator must be in the singlet representation. In the context of
R-symmetry, this was used to classify possible supersymmetry-preserving
deformations in various dimensions \cite{Leigh:1995ep, Green:2010da,
Cordova:2016xhm}. In the example of $\mathcal{N}=4$ super-Yang Mills described
above, the marginal operators are indeed charged under the extra symmetry, and
this explains why they are at finite distance. 

Conversely, if none of the marginal operators are charged under the flavor
symmetry, it cannot be explicitly broken or restored. One would then expect
that in the conformal manifold this point can only be reached asymptotically
and is therefore at infinite distance. This is consistent with a flavor
symmetry emerging as a sector becomes free, namely a higher-spin point. It then
begs the question of the possible existence of other flavor-enhancement points
at infinite distance. According to the second part of the CFT Distance
Conjecture, this cannot be the case, and an emerging flavor symmetry must be
accompanied by a free sector. While in 4d $\mathcal{N}=2$ SCFTs this is
enforced by the superconformal algebra \cite{Dolan:2002zh, Beem:2014zpa,
Cordova:2016emh}, we are not aware of similar arguments in more general cases.
The study of spin-one operators and their behavior in the conformal manifold
therefore seem like an avenue to test the rest of the CFT Distance Conjecture.

\section{Conclusions}\label{sec:conclusions}

A weakly-broken higher-spin symmetry imposes remarkably strong constraints on
the form of the correlators. By studying conformal perturbations close to
higher-spin points, we have harnessed it to show that these points must always
be at infinite Zamolodchikov distance in the conformal manifold, establishing
part of the CFT Distance Conjecture. 

In particular, we have combined two different types of perturbations. First, we
have used conformal perturbation theory to find a criterion which can be used
to decide whether a point is at infinite distance, given in terms of the
behavior of certain OPE coefficients near the point of the conformal manifold
under consideration. Furthermore, we have also utilized the broken Ward
identities appearing close to points where there is a symmetry enhancement.
Through Anselmi's trick, this enabled us to find the leading behavior of the
these coefficients without needing to involve the details of a particular
microscopic description.

\medskip

Combining these two techniques have shown to be powerful as they were
sufficient to show that the first part of the CFT Distance Conjecture is indeed
correct. One can then wonder about the possibility of proving the converse,
namely that infinite-distance points can only be accompanied with a higher-spin
enhancement, and whether the machinery discussed here can be deployed to prove
the rest of the conjecture. One would need to show that the absence of the HS
current at such points lead to an inconsistency, perhaps violating the
assumptions of local unitary CFTs. We expect that the CFT Distance Criterion
could help in this endeavour. We note that efforts have already been spent in
that direction, as it was recently shown using inversion formulae that
singularities of the curvature of the Zamolodchikov metric can only occur in
the presence of higher-spin currents \cite{Balthazar:2022hzb}, or if the CFT
develops a continuum.

Concerning the third part of the conjecture---stating that the anomalous
dimension vanishes exponentially fast with the distance at higher-spin
points---an analysis connected to our results using conservation identities
seems promising. We have indeed shown that this exponential behavior happens
along a given trajectory if $C_{KK\mathcal{O}}^\text{HS}\neq 0$. Therefore, the
third part of the conjecture holds if there is always a marginal operator
$\mathcal{O}$ such that this OPE combination is non-vanishing at the HS point.
Note that the presence of at least one such operators with this property is
enough. The presence of other marginal operators for which
$C_{KK\mathcal{O}}\to 0$ as the HS point is approached would correspond to
non-geodesic trajectories that avoid an exponential behavior, and that are
known to exist in the context of the Swampland Distance Conjecture
\cite{Buratti:2018xjt,Calderon-Infante:2020dhm}. It is furthermore remarkable
that this condition only requires information about the HS point, and one could
hope that this follows from the higher-spin algebra. However, the operators
$K_{\ell-1}$ are not as constrained as the related conserved currents $J_\ell$.
A study of the associated correlators is then more difficult without
considering specific models, and we leave such analyses for future works.

We stress again that, while the evidence for the conjecture has originated from
limiting behaviors of SCFTs, our arguments have not relied on supersymmetry,
and are valid across dimensions. It would however be interesting to investigate
if we can uncover new properties of conformal manifolds by imposing this
additional structure. Part of our analysis might be simplified, and connect
constraints of different spins. There are also examples of conformal manifolds
that are compact \cite{Buican:2014sfa}, and all conformal manifolds of
three-dimensional CFTs have been conjectured to be so as well
\cite{Perlmutter:2020buo}. It would be interesting to check \emph{ad absurdum}
if assuming the presence of a broken Ward identity for the higher-spin symmetry
in those cases leads to inconsistencies. In the same vein, while we have not
assumed the presence of supercharges, we have not tackled the question of
non-supersymmetric conformal manifolds, and only required the presence of
marginal operators regardless of the mechanism protecting them against quantum
corrections. While there have been hints of possible cases
\cite{Giambrone:2021wsm}, their existence is not yet settled.

We have furthermore seen that for spin-one cases, some of the arguments we have
used fail, particularly to show that $C_{JJ\mathcal{O}}^\text{HS}=0$, a key
ingredient of our analysis. Similarly, we could not reach the conclusion
$C_{JK\mathcal{O}}^\text{HS}=0$. This is not a flaw but a feature, since flavor
enhancements are known to happen at both finite and infinite distance in the
conformal manifold. Our analysis leads to simple and intuitive conditions
diagnosing which case occurs. One can then ask whether the condition for having
a flavor-enhancement point at infinite distance would imply that it comes with
additional HS conserved currents. This seems to be a promising scenario for
making progress towards the second part of the CFT Distance Conjecture.

\medskip

Our work falls into the broader context of the holographic Swampland Program,
and provides a step forward in the understanding of the constraints imposed by
quantum gravity through the AdS/CFT correspondence. While we have not discussed
the gravity duals of the local CFTs, our results have implications for theories
of quantum gravity in AdS backgrounds 
\cite{Baume:2020dqd,Perlmutter:2020buo}. A striking feature of our analysis is
that it does not assume that the families of CFTs are holographic, but only
that they have an energy-momentum tensor. This signals that the Swampland
Distance Conjecture could hold beyond weakly-coupled Einstein gravity. On more
general grounds, one may wonder what constraints must be imposed to conformal
theories to address other Swampland conjectures.

Furthermore, our results lend more validity to the physical intuition that
Vasiliev-type gravity---with infinitely many massless HS fields---should be
infinitely far away from Einstein gravity, i.e. they should not be connected by
small deformations \cite{Perlmutter:2020buo}. They are moreover in agreement
with the Emergence proposal---see e.g. references \cite{Harlow:2015lma,
		Heidenreich:2017sim, Grimm:2018cpv, Corvilain:2018lgw, Palti:2019pca,
		Heidenreich:2018kpg, Marchesano:2022axe, Castellano:2022bvr,
Castellano:2023qhp} for works in this direction---which implies that points for
which an infinite number of weakly-coupled fields become massless should always
be at infinite distance. It would also be interesting to connect with the
results that were obtained in references \cite{Stout:2021ubb,Stout:2022phm}, by
further studying how the correlators factorize, particularly when only a
subsector of the theory becomes free. In addition, further exploring conformal
manifolds of local CFTs (possibly with a weakly-coupled gravity dual) and their
possible HS points might shed some light on the fate of the Emergent String
Conjecture \cite{Lee:2019xtm} beyond flat backgrounds, see reference
\cite{Baume:2020dqd} for an analysis of this conjecture applied to examples of
AdS/CFT dual pairs in string theory.

\medskip

Finally, this work is but an example of possible synergies between conformal
theories and the Swampland Program. The Distance Conjectures have been
developed by studying the constraints on field theories imposed by quantum
gravity, but have been applied to conformal manifold independently of this
origin. Other Swampland conjectures might therefore serve as a guide for other
questions on the field-theory side, and motivate particular answers. On the
other hand, conformal techniques have proven to lead to very rigorous results,
and could allow one to establish more thoroughly---or disprove---some of the
more speculative part of the program, and go beyond the current lamppost.

\subsection*{Acknowledgements}

We thank A.~Antunes, S.~Komatsu, M.~Montero, K.~Papadodimas, F.~Tellander,
A.~Uranga, I.~Valenzuela, T.~Weigand and A.~Zhiboedov for helpful discussions
and comments on early versions of the manuscript, as well as E.~Skvortsov for
helpful correspondence. J.C. would like to thank the Department of Physics at
Harvard University and the Instituto de F\'isica Te\'orica in Madrid for
hospitality during early stages of this work. F.B. is supported by the German
Research Foundation through a German-Israeli Project Cooperation (DIP) grant
``Holography and the Swampland'', by the Swiss National Science Foundation
(SNSF), grant number P400P2\_194341, and in part by the Deutsche
Forschungsgemeinschaft under Germany’s Excellence Strategy EXC 2121 Quantum
Universe 390833306. The work by J.C. is partially supported by the Spanish
Agencia Estatal de Investigacion through the grant “IFT Centro de Excelencia
Severo Ochoa CEX2020-001007-S, the grants PGC2018-095976-B-C21 and
PID2021-123017NB-I00, funded by MCIN/AEI/10.13039/ 501100011033, by ERDF A way
of making Europe, and the FPU grant no. FPU17/04181 from the Spanish Ministry
of Education.

\appendix

\section{Current-Conservation Conditions} \label{sec:appendix-A}

In Section \ref{sec:HS-at-infinite-distance}, we discuss the various
constraints the higher-spin algebra imposes on correlators. In particular, the
Ward identity $\partial\cdot J_\ell = 0$ can be used to relate some of the OPE
coefficients of a given three-point function involving $J_\ell$. As pointed out
in references \cite{Maldacena:2011jn, Zhiboedov:2012bm}, taking the divergence
of the three-point functions can be rephrased in terms of a differential
operator acting in the space of the building blocks constituting the conformal
structures. For the correlators we are considering, there are only three
building blocks for the conformal structures defined in equations
\eqref{conformal-structure} and \eqref{conformal-structure-tilde}: $H_{12}$,
$V_1$ and $V_2$ (their expressions as function of the coordinates and
polarization vectors are shown in equation \eqref{building-blocks}).

In all generality, the expression of the differential operator for
type-$(\ell_1,\ell_2,\ell_3)$ correlators depends on all six building blocks,
and was given in reference \cite{Zhiboedov:2012bm}, whose conventions we
follow. For our purpose, we will always consider three-point functions
involving a marginal operator, $\ell_3=0$, which simplifies considerably the
expressions we will have to deal with. We will successively treat the two cases
at hand, namely $\langle J_\ell J_\ell\mathcal{O}\rangle$ and $\langle J_\ell
K_{\ell-1}\mathcal{O}\rangle$. For clarity, we reproduce here the conformal
structure in both cases, using that $J_\ell$ has twist $d-2$, while
$K_{\ell-1}$ and $\mathcal{O}$ have twist $d$:
\begin{equation}\label{conformal-structure-numerator}
	\begin{aligned}
			\Theta_n &= \frac{\mathcal{P}_n}{|x_{12}|^{d-4}|x_{13}|^d|x_{23}|^d}\,,\qquad& \mathcal{P}_n &= H_{12}^{\ell-n}(V_1V_2)^n\,,\\
			\widetilde{\Theta}_n &= \frac{\widetilde{\mathcal{P}}_n}{|x_{12}|^{d-2}|x_{13}|^{d-2}|x_{23}|^{d+2}}\,,\qquad& \widetilde{\mathcal{P}}_n &= H_{12}^{\ell-n-1}V_1^{n+1}V_2^n\,.
	\end{aligned}
\end{equation}
We have denoted the numerators $\mathcal{P}$ and $\tilde{\mathcal{P}}$, as
these are the quantities on which the differential operator discussed above is
applied.

\subsection*{Current Conservation for $\langle J_\ell J_{\ell}\mathcal{O}\rangle$}

In this case, the conservation equation leads to the following constraint:
\begin{equation}
		\langle J_\ell(x_1,\xi_1) (\partial\cdot J_\ell)(x_2,\xi_2)\mathcal{O}(x_3)\rangle=0\quad\Leftrightarrow\quad
		\mathcal{D}_2\,\langle J_\ell(x_1,\xi_1) J_\ell(x_2,\xi_2)\mathcal{O}(x_3)\rangle =0\,.
\end{equation}
where the differential operator, understood as acting on the second current,
reduces from the original expression \cite{Zhiboedov:2012bm} to: 
\begin{equation}
    \mathcal D_2 = \frac{\mathcal D_{2}^{(3)}}{d-2} + \mathcal D_{2}^{(2)} + (d-2) \mathcal D_{2}^{(1)} -2 \left( \frac{\delta \mathcal D_{2}^{(2)}}{d-2} + \delta \mathcal D_{2}^{(1)} \right) \, ,
\end{equation}
with
\begin{equation} \label{suboperators}
\begin{split}
    \mathcal D_{2}^{(1)} &= V_1 \,\partial_{H_{12}}  \, , \\
    \mathcal D_{2}^{(2)} &= 2 H_{12} V_1 \, \partial_{H_{12}}^{2} + V_1 \, \partial_{V_1} \partial_{V_2} + \left( H_{12} + 2 V_1 V_2 \right) \, \partial_{V_2} \partial_{H_{12}}\, , \\
    \mathcal D_{2}^{(3)} &= 2H_{12}\left( H_{12} + 2 V_1 V_2 \right) \, \partial_{V_2} \partial_{H_{12}}^{2} + \left( H_{12} + 2 V_1 V_2 \right) \, \partial_{V_1} \partial_{V_2}^{2}\, , \\
    \delta \mathcal D_{2}^{(1)} &= \partial_{V_2} - V_1 \, \partial_{H_{12}} \, , \\
    \delta \mathcal D_{2}^{(2)} &= V_2\, \partial_{V_2}^{2} + 2 H_{12} \, \partial_{V_2} \partial_{H_{12}} - 2 V_1 H_{12} \, \partial_{H_{12}}^{2} \, . \\
\end{split}
\end{equation}
Note that we can safely ignore all terms involving building blocks that do not
appear in the type-$(\ell,\ell,0)$ conformal structures. Furthermore, in
reference \cite{Zhiboedov:2012bm}, the operator is defined as acting on the
first operator. To get to the above equation, we have exchanged the indices
$1\leftrightarrow2$.

As explained in Section \ref{sec:HS-at-infinite-distance}, this operator can
then be applied to each of the conformal structures, $\Theta_n$. The
differential operator is taking care of the contribution from the denominator
by design, and we find the following relation for the numerator of a given
structure:
\begin{equation}
\begin{split}
    (d-2) \mathcal{D}_2 \mathcal P_n &= n^2 (n-1) \, \widetilde{\mathcal P}_{n-2} \\
    &+ n \left[ d(\ell -2) + 2 ( 3 + \ell^2 - 2\ell (n+2) +2n^2 +n ) \right] \, \widetilde{\mathcal P}_{n-1} \\
    &+ (\ell-n) (d+2n) (d-4 + 2\ell -2n) \, \widetilde{\mathcal P}_{n} \, .
\end{split}
\end{equation}
As expected, all dependence on the building blocks can be written in terms of
the numerator of the type-$(\ell,\ell-1,0)$ conformal structures
$\widetilde{\Theta}_n$, see equation \eqref{conformal-structure-numerator}. We
can then plug in those relations in the numerator of the three-point function
$\langle J_\ell J_{\ell}\mathcal{O}\rangle$:
\begin{equation}
		\mathcal P = \sum_{n=0}^{\ell} C_{JJ\mathcal{O}}^{\,n} \mathcal P_{n} \,.
\end{equation}

After shifting the sums and a bit of algebra to simplify the expression,
everything can be rewritten as a single sum:
\begin{equation}
\begin{split}
	(d-2) \mathcal{D}_2 \mathcal P &= \sum_{n=0}^{\ell-1} \bigg{[} (n+2)^2 (n+1) \, C_{JJ\mathcal{O}}^{n+2} \\
	&+ (n+1) \left[ d(\ell -2) + 2 ( 6 + \ell^2 - 2\ell (n+3) + 2n^2 + 5n ) \right]\, C_{JJ\mathcal{O}}^{n+1} \\
	&+ (\ell-n) (d+2n) (d-4 + 2\ell -2n) \, C_{JJ\mathcal{O}}^n \bigg{]} \, \widetilde{\mathcal P}_{n} \, ,
\end{split}
\end{equation}
where we have defined $C_{JJ\mathcal{O}}^{\ell+1} = 0$. Note that this is a
boundary condition, as the OPE coefficient $C_{JJ\mathcal{O}}^{n}$ have a
physical meaning only for $0\leq n\leq\ell$. The current conservation condition
then imposes that the prefactors in front of each $\widetilde{\mathcal P}_{n}$
must all vanish, leading to a system of constraints:
\begin{equation} \label{system-of-equations}
    A(n) \, C_{JJ\mathcal{O}}^{n+2} + B(n) \, C_{JJ\mathcal{O}}^{n+1} + D(n) \, C_{JJ\mathcal{O}}^{n} = 0 \, , \quad n=0,\dots,\ell-1 \,,
\end{equation}
where we have defined
\begin{equation}
\begin{split}
  A(n) &= (n+2)^2 (n+1) \, , \\
  B(n) &= (n+1) \left[ d(\ell -2) + 2 ( 6 + \ell^2 - 2\ell (n+3) + 2n^2 + 5n ) \right] \, , \\
  D(n) &= (\ell-n) (d+2n) (d-4 + 2\ell -2n) \,,
\end{split}
\end{equation}
and one has to keep in mind the boundary condition:
\begin{equation}
    C_{JJ\mathcal{O}}^{\ell+1}=0 \, .
\end{equation}
As discussed in the main text, this is a homogeneous system of $\ell$ equations
for $\ell+1$ variables. For given $d>2$ and $\ell\geq 0$, one can check that
$D(n) \neq 0$ for $0 \leq n \leq \ell -1$, and we can rewrite this as:
\begin{equation}\label{CJJO-recursion-relation}
	C_{JJ\mathcal{O}}^{n} = - \frac{1}{D(n)} \left( A(n) \, C_{JJ\mathcal{O}}^{n+2} + B(n) \, C_{JJ\mathcal{O}}^{n+1} \right) \, , \quad n=0,\dots,\ell-1 \,.
\end{equation}

In this form, we clearly see that this is a recursion relation in the space of
OPE coefficients $C_{JJ\mathcal{O}}^n$. Using $C^{\ell}_{JJ\mathcal{O}}$ and
$C_{JJ\mathcal{O}}^{\ell+1}=0$ as initial conditions and starting from
$n=\ell-1$, we can find any of the OPE coefficients as being proportional to
$C^{\ell}_{JJ\mathcal{O}}$, and we indeed find the result given in equation
\eqref{eq:recursion-relation}.

Notice that since $A(n) \neq 0$ for $0 \leq n \leq \ell -1$, we could instead
have started with $n=0$ and solved the system in ascending order. However, in
this way $C_{JJ\mathcal{O}}^{\ell+1}=0$ cannot be used as an initial condition.
This will lead to solutions in terms of $C_{JJ\mathcal{O}}^{0}$ and
$C_{JJ\mathcal{O}}^{1}$, plus the constraint imposed by the boundary condition
$C_{JJ\mathcal{O}}^{\ell+1}=0$. As it is not guaranteed that the last
constraint is non-trivial, we find it more convenient to write the solution in
terms of $C_{JJ\mathcal{O}}^n$.
 
\medskip
As advertised in Section \ref{sec:HS-at-infinite-distance}, given the current
conservation condition for $\langle J_\ell J_\ell \mathcal{O} \rangle$, we have
shown that we can always write all the OPE coefficients $C_{JJ\mathcal{O}}^n$
in terms of $C_{JJ\mathcal{O}}^\ell$ as in equation
\eqref{eq:recursion-relation}. In other words, this condition reduces the
number of allowed conformal structures from $\ell+1$ to only one. For arbitrary
dimension and spin, we have not found a closed-form solution of the recursion
relation. However, it can be easily solved for fixed values programmatically,
see equations \eqref{vn-spin-1} and \eqref{vn-spin-4} for $\ell=1,4$.

\subsection*{Current Conservation for $\langle J_\ell K_{\ell-1}\mathcal{O}\rangle$}

Imposing the divergence of the higher-spin current $J_\ell$ in the case of the
three-point function $\left< J_\ell K_{\ell-1} \mathcal O\right>$ proceeds
\emph{mutatis mutandis} from that of $\langle J_\ell
J_{\ell}\mathcal{O}\rangle$. The divergence is again translated into a
differentiation in the space of building blocks:
\begin{equation}
		\langle (\partial\cdot J_\ell)(x_1,\xi_1) K_{\ell-1}(x_2,\xi_2)\mathcal{O}(x_3)\rangle=0\quad\Leftrightarrow\quad
		\mathcal{D}_1\,\langle J_\ell(x_1,\xi_1) K_{\ell-1}(x_2,\xi_2)\mathcal{O}(x_3)\rangle =0
\end{equation}
The differential operator applied to the first operator is now simpler than in
the previous case, and given by \cite{Zhiboedov:2012bm}:
\begin{equation}\label{differential-operator-JKO}
    \mathcal D_1 = \frac{\mathcal D_{1}^{(3)}}{d-2} + \mathcal D_{1}^{(2)} + (d-2) \mathcal D_{1}^{(1)} \, ,
\end{equation}
where we have defined each component as:
\begin{equation} \label{suboperators2}
\begin{split}
    \mathcal D_{1}^{(1)} &= V_2 \,\partial_{H_{12}} \,,\\
    \mathcal D_{1}^{(2)} &= 2 H_{12} V_2 \, \partial_{H_{12}}^{2} + V_2 \, \partial_{V_1} \partial_{V_2} + \left( H_{12} + 2 V_1 V_2 \right) \, \partial_{V_1} \partial_{H_{12}}\, , \\
    \mathcal D_{1}^{(3)} &= 2H_{12}\left( H_{12} + 2 V_1 V_2 \right) \, \partial_{V_1} \partial_{H_{12}}^{2} + \left( H_{12} + 2 V_1 V_2 \right) \, \partial_{V_1}^{2} \partial_{V_2} \, . \\
\end{split}
\end{equation}

Once again, following the procedure explained in Section
\ref{sec:HS-at-infinite-distance}, this operator is applied to the numerator of
the correlator. The numerators $\widetilde{\mathcal{P}}_{n}$ of the
type-$(\ell,\ell-1,0)$ conformal structure $\widetilde{\Theta}_n$ then
satisfies the following relation:
\begin{equation}
\begin{split}
   (d-2) \mathcal{D}_1 \widetilde{\mathcal P}_n &= (n+1) n^2 \, \mathcal P_{n-1} \\
   &+ (n+1) \left[ (\ell-1)(d+2\ell-6) - (4\ell-6)n + 4n^2 \right] \, \mathcal P_{n} \\
   &+ (\ell -1 -n) (d+2n) (d-6+2\ell-2n) \, \mathcal P_{n+1} \, .
\end{split}
\end{equation}
As expected, all dependence on the building blocks can be written in terms of
the numerator of type-$(\ell-1,\ell-1,0)$ conformal structures $P_n$, see
equation \eqref{conformal-structure-numerator}.\footnote{Note that in equation
\eqref{conformal-structure-numerator} the conformal structures are given for
spin $\ell$, and one therefore needs to substitute $\ell\to\ell-1$.} Plugging
these relations into the numerator of the full three-point function $\left<
J_\ell K_{\ell-1} \mathcal O\right>$:
\begin{equation}
   \widetilde{\mathcal P} = \sum_{n=0}^{\ell-1} C_{JK\mathcal{O}}^n \widetilde{\mathcal P}_{n} \, ,
\end{equation}
we then have
\begin{equation}
\begin{split}
   (d-2) \mathcal{D}_1 \widetilde{\mathcal P} &= \sum_{n=0}^{\ell-1} (n+1) n^2 \, C_{JK\mathcal{O}}^n \mathcal P_{n-1} \\
   &+ \sum_{n=0}^{\ell-1} (n+1) \left[ (\ell-1)(d+2\ell-6) - (4\ell-6)n + 4n^2 \right] \, C_{JK\mathcal{O}}^n \mathcal P_{n} \\
   &+ \sum_{n=0}^{\ell-1} (\ell -1 -n) (d+2n) (d-6+2\ell-2n) \, C_{JK\mathcal{O}}^n \mathcal P_{n+1} \, .
\end{split}
\end{equation}
Shifting the sums, we can rewrite the sums in terms of summands containing only
$\mathcal{P}_n$ if we impose
$C_{JK\mathcal{O}}^{-1}=0=C_{JK\mathcal{O}}^{\ell}$. Note that this is not a
constraint on the OPE coefficients, as for type-$(\ell,\ell-1,0)$ correlators
$C_{JK\mathcal{O}}^n$ has a physical meaning only for $0\leq n\leq\ell-1$.
After a few algebraic manipulations, we find:
\begin{equation}
\begin{split}
    (d-2) \mathcal{D}_1 \widetilde{\mathcal P} &= \sum_{n=0}^{\ell-1} \bigg{[} (n+2)(n+1)^2 \, C_{JK\mathcal{O}}^{n+1} \\
    &+ (n+1) \left[ (\ell-1)(d+2\ell-6) - (4\ell-6)n + 4n^2 \right] \, C_{JK\mathcal{O}}^n  \\
    &+ (\ell - n) (d-2+2n) (d-4+2\ell-2n) \, C_{JK\mathcal{O}}^{n-1} \bigg{]} \mathcal P_{n} \, .
\end{split}
\end{equation}

The current-conservation condition then imposes that the prefactors of each
$\mathcal P_{n}$ to vanish, and we once more obtain:
\begin{equation}
    \tilde A(n) \, C_{JK\mathcal{O}}^{n+1} + \tilde B(n) \, C_{JK\mathcal{O}}^{n} + \tilde D(n) \, C_{JK\mathcal{O}}^{n-1} = 0 \, , \quad n=0,\dots,\ell-1 \,,
\end{equation}
where we have defined
\begin{equation}
\begin{split}
    \tilde A(n) &= (n+2)(n+1)^2 \, , \\
    \tilde B(n) &= (n+1) \left[ (\ell-1)(d+2\ell-6) - (4\ell-6)n + 4n^2 \right] \, , \\
    \tilde D(n) &= (\ell - n) (d-2+2n) (d-4+2\ell-2n) \, ,
\end{split}
\end{equation}
and we impose the following boundary conditions:
\begin{equation}
    C_{JK\mathcal{O}}^{-1}=0=C_{JK\mathcal{O}}^{\ell} \, .
\end{equation}
Taking into account the constraints, this is a homogeneous system of $\ell$
equations for $\ell$ variables. If it is closed, all the OPE coefficients must
vanish. However, in practice we find that only $\ell-1$ equations are
independent, and the OPE coefficients $C_{JK\mathcal{O}}^{n}$ are all related
to each other. Dealing with this system of equations for arbitrary $\ell$ is
however tedious because the spin controls the number of equations and
variables, but it can easily be solved for fixed values of the dimension and
spin through computational means. 

Despite this, in what follows we show that in the worst-case scenario, this
system of equations always allows us to fix all OPE coefficients in terms of
$C_{JK\mathcal{O}}^{\ell-1}$. Conversely, in the best-case scenario the system
is closed and we further have $C_{JK\mathcal{O}}^{\ell-1}=0$---we have checked
that, unfortunately, this only happens in a few sporadic cases.

\medskip

Given $d>2$ and $\ell\geq 0$, one can check that $\tilde D(n) \neq 0$ for $0
\leq n \leq \ell -1$. Therefore, we can write this system of equations as
\begin{equation} \label{system-of-equations-2}
     C_{JK\mathcal{O}}^{n-1} = - \frac{1}{\tilde D(n)} \left( \tilde A(n) \, C_{JK\mathcal{O}}^{n+1} + \tilde B(n) \, C_{JK\mathcal{O}}^{n} \right) \, , \quad n=0,\dots,\ell-1 \,.
\end{equation}
In this form, we clearly see that this is a recursion relation in the space of
OPE coefficients $C_{JK\mathcal{O}}^n$. In particular, we can successively
solve it for descending values of $n$. Using $C^{\ell-1}_{JK\mathcal{O}}$ and
$C_{JK\mathcal{O}}^{\ell}=0$ as initial conditions, it allows us to write all
the OPE coefficients in terms of $C^{\ell-1}_{JK\mathcal{O}}$. In fact, since
this is a set of linear equations, the result will take the form of equation
\eqref{eq:recursion-relation-2}. 

Unlike for the case of $\left< J_\ell J_\ell \mathcal O \right>$, this is not
the end of the story. Notice that this result is also valid for
$C_{JK\mathcal{O}}^{-1}$, which by definition is vanishing. This is, we have an
extra constraint
\begin{equation}
    C_{JK\mathcal{O}}^{-1} = \tilde{v}^{-1} C_{JK\mathcal{O}}^{\ell-1} = 0 \, .
\end{equation}
This has two possible outputs: either $\tilde{v}^{-1}\neq 0$ and this
constraint further imposes that $C_{JK\mathcal{O}}^{\ell-1}$ must vanish, or
$\tilde{v}^{-1}= 0$ and the constraint is trivial. The first option corresponds
to the system of linear equations being closed, since we have $\ell$ equations
for $\ell$ variables. An easy way of distinguishing these two cases for some
fixed $\ell$ is to compute the determinant of the matrix corresponding to the
system of equations in \eqref{system-of-equations-2} and check whether it
vanishes. As mentioned in the main text, we checked that it indeed does in any
dimension for $2 \leq \ell \leq 10$. In any case, we either find
$C_{JK\mathcal{O}}^{n}=0$, which is one of the ingredient of the proof in
Section \ref{sec:HS-at-infinite-distance}, or we find the result shown in
equation \eqref{eq:recursion-relation-2}, and need to apply the integrated Ward
identity as explained in the same section.

Note that, since $A(n) \neq 0$ for $0 \leq n \leq \ell -1$, we could have also
solved the recursion relation \eqref{system-of-equations} in ascending order,
which would give all OPE coefficients in terms of $C_{JK\mathcal{O}}^{0}$. In
this case, we would find an extra constraint, $C_{JK\mathcal{O}}^{\ell}=0$.
This is in contrast with the case of $\left< J_\ell J_\ell \mathcal O \right>$,
where solving the recursion relation in descending order avoided the appearance
of an extra constraint.

\medskip

To summarize, we have shown that the current conservation condition applied to
$\left< J_\ell K_{\ell-1} \mathcal O \right>$, allows us to always write the
OPE coefficients $C_{JK\mathcal{O}}^n$ in terms of $C_{JK\mathcal{O}}^{\ell-1}$
as described in equation \eqref{eq:recursion-relation-2}. In other words, this
condition reduces (at least) the $\ell$ allowed conformal structures to only
one.

\subsection*{Current Conservation for $\langle J_\ell K_{3}\mathcal{O}\rangle$}

Another type of correlators that appear in the main text are those of type
$(\ell,3,0)$. As we will only need to deal with $\ell=1,2,3$  we will
not repeat the whole analysis for arbitrary $\ell$ as we did so far in 
this appendix, but rather focus on those special cases. The numerators of those correlators then takes the form
\begin{equation}\label{numerator-P-hat}
		\widehat{\mathcal{P}}_{\ell} = \sum_{n=0}^\ell C^n_{JK_3\mathcal{O}} H_{12}^{\ell-n}V_1^nV_2^{n-\ell+3}\,,\qquad 0<\ell\leq3\,.
\end{equation}
As the differential operator enforcing the conservation condition depends only
on the difference in the twist of the two non-conserved operators, we can use
the one given in equation \eqref{differential-operator-JKO}. As we focus on low
spins, finding the constraints by expanding the result in terms of the building
blocks is straightforward, and we again find that there is a unique independent
OPE coefficient:
\begin{equation}\label{conservation-result-jk3O}
\begin{split}
  \ell=1 &:\qquad C_{JK_3\mathcal{O}}^0 =- \frac{3}{d-2}C_{JK_3\mathcal{O}}^1\,;\\
		\ell=2 &:\qquad C_{JK_3\mathcal{O}}^0 = \frac{6}{d(d-2)}C_{JK_3\mathcal{O}}^2\, ,\qquad C_{JK_3\mathcal{O}}^2 = \frac{6}{d-2}C_{JK_3\mathcal{O}}^2\,;\\
		\ell=3 &:\qquad C_{JK_3\mathcal{O}}^0 = \frac{-6C_{JK_3\mathcal{O}}^3}{d(d^2-4)}\, ,\qquad C_{JK_3\mathcal{O}}^1 = \frac{18C_{JK_3\mathcal{O}}^3}{d(d-2)}\, ,\qquad C_{JK_3\mathcal{O}}^2 = \frac{-9C_{JK_3\mathcal{O}}^3}{d-2}\,.
\end{split}
\end{equation}

\subsection*{Current Conservation for $\langle J_2 K_{\ell}\mathcal{O}\rangle$}

Correlators of type $(2,\ell,0)$ also appear in the charge-conservation identity discussed in Section \ref{sec:weakly-broken}. Here we will focus in the cases $\ell=1,\dots,6$, as needed for the analysis in the main text. The numerators of those correlators then take the form
\begin{equation}\label{numerator-p-prime}
		\mathcal{P}^{\prime}_{\ell} = \sum_{n=0}^{\min(\ell,2)} C^n_{J_2 K \mathcal{O}} H_{12}^{n}V_1^{2-n}V_2^{\ell-n} \, .
\end{equation}
The differential operator is again the one given in equation \eqref{differential-operator-JKO}. As we focus on low
spins, finding the constraints by expanding the result in terms of the building
blocks is straightforward, and we again find that there is a unique independent
OPE coefficient:
\begin{equation} \label{conservation-result-j2kO}
\ell=1,\cdots,6 :\qquad C_{J_2 K\mathcal{O}}^0 = -\frac{d-2}{2\ell}C_{J_2 K\mathcal{O}}^1\, ,\qquad C_{J_2 K\mathcal{O}}^2 = -\frac{\ell-1}{2d}C_{J_2 K\mathcal{O}}^1\,.
\end{equation}

\section{Surface Integrals and Ward Identities}\label{app:surface-integrals}

The previous appendix dealt with the current conservation condition applied to
three-point functions, and the constraints it imposed on the OPE coefficients.
We have found that in the two cases of interest, it reduces the number
parameters to (at most) one. In Section \ref{sec:HS-at-infinite-distance}, we
further argue that the remaining coefficient must vanish at the higher-spin
point using the conserved charges obtained by integrating conserved currents
over a codimension-one surface $\Sigma$. This requires a choice of surface and
polarization vector so that the integral over the conformal structure is
non-trivial. We now show that there is always a choice leading to the desired
outcome. 

To do so, we will need the expression of the three building blocks in terms of
the spacetime coordinates and polarization vectors:
\begin{equation} \label{building-blocks}
\begin{aligned}
    H_{12} &= \frac{(\xi_1 \cdot \xi_2) |x_{12}|^{2} - 2 (\xi_1 \cdot x_{12})(\xi_2 \cdot x_{12})}{|x_{12}|^4} \,, \\
    V_1 &= \frac{(\xi_1 \cdot x_{13})}{|x_{13}|^2} - \frac{(\xi_1 \cdot x_{12})}{|x_{12}|^2} \,, \\
    V_2 &= \frac{(\xi_2 \cdot x_{21})}{|x_{21}|^2} - \frac{(\xi_2 \cdot x_{23})}{|x_{23}|^2} \,.
\end{aligned}
\end{equation}
In both cases of interest, and following reference \cite{Osborn:1993cr}, we
consider the theory in Euclidean spacetime and fix the position of the three
operators to be at:
\begin{equation}\label{integral-point-choice}
    x_{1} = (0,\mathbf x) \, , \quad x_{2} = (\frac{1}{2} y,\mathbf 0) \, , \quad x_{3} = (-\frac{1}{2} y,\mathbf 0) \, .
\end{equation}
This simplifies the expressions of the conformal structures. For instance, we
have the relations:
\begin{equation} \label{modulus}
	|x_{12}|^2 = \mathbf x^2 + \frac{1}{4} y^2\, , \quad |x_{13}|^2 = \mathbf x^2 + \frac{1}{4} y^2 \, , \quad |x_{23}|^2 = y^2\,,
\end{equation}

We then need a codimension-one surface over which we integrate the first
higher-spin conserved current $J_\ell$. A choice that reveals itself to be very
convenient is the surface defined by $x_{1} = (0,\mathbf x)$. It corresponds to
picking the $J_{\ell}^{1 \nu_1 \cdots \nu_{\ell-1} }$ components of the current
in the correlator. It is then easier to demand all indices to be in the same
direction, and we choose the first polarization vector to be $\xi_1 =(1,\mathbf
0)$. The second polarization vector remains unfixed, and for our purpose the
following choice will lead us to the desired outcome: 
\begin{equation}\label{polarization-choice}
		\xi_1 =\xi_{2}=(1,\mathbf 0)\,,
\end{equation}
which in turn allows us to rewrite the building blocks as:
\begin{equation}
	H_{12} = \frac{\mathbf x^2 - \frac{1}{4} y^2}{(\mathbf x^2 + \frac{1}{4} y^2)^2} \, \quad
    V_1 = \frac{y}{\mathbf x^2 + \frac{1}{4} y^2} \, , \quad
    V_2 = - \frac{\mathbf x^2 - \frac{1}{4} y^2}{y(\mathbf x^2 + \frac{1}{4} y^2)} \, .
\end{equation}
Interestingly, this choice implies an additional relation between the three
buildining blocks that will considerably simplify the form of the conformal
structure, and allow us to perform the integrals exactly:
\begin{equation}\label{VV=-H}
    V_1 V_2 = -H_{12} \, .
\end{equation}

With the definitions above, we are now ready to consider the surface integral
for both relevant three-point functions necessary to use the integrated Ward
identity.

\subsection*{Integrals of $\langle J_\ell J_\ell \mathcal{O}\rangle$ Structures}

In the case of surface integrals of the three-point functions $\langle J_\ell
J_\ell \mathcal{O}\rangle$, the conformal structures are given by equation
\eqref{conformal-structure}. For the choice of polarizations given above and
the relation \eqref{VV=-H}, all the conformal structure take the same form up
to a sign:
\begin{equation}
    \Theta_{n} = \frac{(-1)^{n} \left( \mathbf x^2 - \frac{1}{4} y^2 \right)^{\ell}}{y^{d} \left( \mathbf x^2 + \frac{1}{4} y^2 \right)^{d-2+2\ell}} \,, 
\end{equation}
We are therefore looking for an expression for the integral over the surface
$\Sigma$ defined by $x_1=(0,\mathbf{x})$ and for the choice of polarization
given in equation \eqref{polarization-choice}:
\begin{equation}
		I_{JJ\mathcal{O}}(\ell) =\sum_{n=0}^{\ell}\int_\Sigma d^{d-1}\mathbf{x}\, v^n\Theta_n = 
		\frac{\sum_{n=0}^{\ell} (-1)^{n} v^{n}}{y^{d}} \int_{\mathbb{R}^{d-1}} d^{d-1}\mathbf x \, \frac{\left( \mathbf x^2 - \frac{1}{4} y^2 \right)^{\ell}}{\left( \mathbf x^2 + \frac{1}{4} y^2 \right)^{d-2+2\ell}} \, ,
\end{equation}
where the coefficients $v^n$ are those found by solving the recursion relation
\eqref{CJJO-recursion-relation} imposing the current-conservation condition
discussed in Section \ref{sec:charge-conservation}. As mentioned there, while
we do not have a closed expression for the coefficients $v^n$ for arbitrary
spin, we have checked that the prefactor does not vanish,
$\sum_{n=0}^\ell(-1)^n v^n\neq0$, for $\ell=1,2,\dots,10$.

This integral can be then performed straightforwardly by noting that the
integration variable only appear through its norm. A change of variable
$\mathbf{x}=y\mathbf{\tilde{x}}$ followed by going to spherical coordinates
leads us to:
\begin{equation}
		I_{JJ\mathcal{O}}(\ell) = \frac{\sum_{n=0}^{\ell} (-1)^{n} v^{n}}{y^{2d-3+2\ell}} \, S_{d-2} \int_{0}^{\infty} dr \, \, \frac{r^{d-2} \left(r^2 - \frac{1}{4} \right)^{\ell}}{\left( r^2 + \frac{1}{4} \right)^{d-2+2\ell}} \,,
\end{equation}
where $S_{d-2}$ is the volume of the $(d-2)$-sphere. Using the integral
representation of hypergeometric functions, we finally have the closed-form
expression:
\begin{equation}
\begin{split}
    \int_{0}^{\infty} dr \, \, \frac{r^{d-2} \left(r^2 - \frac{1}{4} \right)^{\ell}}{\left( r^2 + \frac{1}{4} \right)^{d-2+2\ell}} &= 
    (-1)^\ell\, 2^{d-2+2(\ell-1)}\frac{\Gamma(\frac{d-1}{2})\Gamma(\frac{d-3+4\ell}{2})}{\Gamma(d-2+2\ell)}\,{}_2F_1\left(\frac{d-1}{2}, -\ell, -\frac{d -5+4\ell}{2}; -1\right)
\end{split}
\end{equation}

We conclude that for $d>2$ and $\ell>1$, the surface integral
$I_{JJ\mathcal{O}}$ does not generically vanish. For instance, in the case of
spin-four currents, the surface integrals reduces to:
\begin{equation}
	I_{JJ\mathcal{O}}(\ell=4)
    = S_{d-2} \, \frac{3(d+19)\sqrt{\pi} \, \Gamma\left( \frac{d-1}{2} \right)}{\Gamma\left( \frac{d}{2}+3 \right)}\frac{\sum_{n=0}^{4} (-1)^{n} v^{n}}{y^{2d+5}} \, .
\end{equation}

In Section \ref{sec:HS-at-infinite-distance}, this enables us to show that
three-point function $\left<J_4 J_4\mathcal{O}\right>$ is trivial, an important
step in our proof. Note however that for spin-one currents, we cannot arrive at
this conclusion with the choice of polarizations we have made, as for any
dimension,
\begin{equation}
		I_{JJ\mathcal{O}}(\ell=1) = 0\,.
\end{equation}
In fact, this integral always vanishes for the spin-one case. Repeating the
above computation for the polarizations $\xi_1=(1,\mathbf{0})$ and
$\xi_2=(0,\bm{\xi})$, we are led to
\begin{equation}
		\sum_{n=0}^\ell\int_\Sigma d^{d-1}\mathbf x \, v^{n} \Theta_{n} = \frac{\sum_{n=0}^{\ell} (-1)^{n} v^{n}}{y^{d-1}} \int_{\mathbb{R}^{d-1}} d^{d-1}\mathbf x \, \frac{\left( \pmb \xi \cdot \mathbf x \right)}{\left( \mathbf x^2 + \frac{1}{4} y^2 \right)^{d}} = 0 \,,
\end{equation}
where we have used that as the integrand is odd, for our choice of surface the
integral is trivial. As a generic polarization can be written as a combination
of the two choices above and the integrals are linear with respect to the
polarization, this surface integral vanishes for $\ell=1$ for any choice of
$\xi_2$.

\subsection*{Integrals of $\langle J_\ell K_{\ell-1}\mathcal{O}\rangle$ Structures}

We can repeat the procedure used above for the three-point function $\langle
J_\ell K_{\ell-1}\mathcal{O}\rangle$. Placing the spacetime points as in
equation \eqref{integral-point-choice}, the type-$(\ell,\ell-1,0)$ conformal
structures (see equation \eqref{conformal-structure-tilde} for a definition),
we obtain:
\begin{equation}
    \widetilde{\Theta}_{n} = \frac{ (-1)^{n} \left( \mathbf x^2 - \frac{1}{4} y^2 \right)^{\ell-1}}{y^{d+1} \left( \mathbf x^2 + \frac{1}{4} y^2 \right)^{d-3+2\ell}} \, ,
\end{equation}
For the choice of surface $\Sigma$ defined by $x_1=(0,\mathbf{x})$, and the
polarization given in equation \eqref{polarization-choice}, we want to
calculate:
\begin{equation}
		I_{JK\mathcal{O}}(\ell) =\sum_{n=0}^{\ell}\int_\Sigma d^{d-1}\mathbf{x}\, \tilde{v}^n\widetilde{\Theta}_n = 
		\frac{\sum_{n=0}^{\ell-1} (-1)^{n} \tilde{v}^{n}}{y^{d+1}} \int d^{d-1}\mathbf x \, \frac{\left( \mathbf x^2 - \frac{1}{4} y^2 \right)^{\ell-1}}{ \left( \mathbf x^2 + \frac{1}{4} y^2 \right)^{d-3+2\ell}} \, .
\end{equation}

The coefficients $\tilde{v}^n$ are given by the recursion relation
\eqref{system-of-equations-2} imposed by the current conservation condition. We
have have checked that the prefactor is non-zero in any dimension for
$\ell=1,2,\dots,10$: $\sum_{n=0}^{\ell-1}(-1)^n\tilde{v}^n\neq0$ . Turning to
the integral itself, using the same change of variables as above, we obtain:
\begin{equation}
	I_{JK\mathcal{O}}(\ell) = \frac{\sum_{n=0}^{\ell-1} (-1)^{n} \tilde{v}^{n}}{y^{d-2+2\ell}} \, S_{d-2} \int_{0}^{\infty} dr \, \, \frac{r^{d-2} \left(r^2 - \frac{1}{4} \right)^{\ell-1}}{\left( r^2 + \frac{1}{4} \right)^{d-3+2\ell}} \, .
\end{equation}
Using the integral representation of hypergeometric functions, we finally find
that:
\begin{equation}
\begin{split}
    \int_{0}^{\infty} dr \, \, \frac{r^{d-2} \left(r^2 - \frac{1}{4} \right)^{\ell-1}}{\left( r^2 + \frac{1}{4} \right)^{d-3+2\ell}} &= 
    (-1)^{\ell+1}\, 2^{d-2+2(\ell-1)}\frac{\Gamma(\frac{d-1}{2})\Gamma(\frac{d-5+4\ell}{2})}{\Gamma(d-3+2\ell)}\,{}_2F_1\left(\frac{d-1}{2}, -\ell+1, -\frac{d -7+4\ell}{2}; -1\right)
\end{split}
\end{equation}
and $I_{JK\mathcal{O}}(\ell)$ never vanishes for any value of the dimension or
the spin. For $\ell=1,4$, it takes the simple form:
\begin{align}
    I_{JK\mathcal{O}}(\ell=1) &= \frac{1}{y^{2d}} \, S_{d-2}\frac{\sqrt{\pi}\, \Gamma(\frac{d-1}{2})}{\Gamma(\frac{d}{2})} = \frac{1}{y^{2d}} \, S_{d-1} \,, \\
    I_{JK\mathcal{O}}(\ell=4) &= \frac{\sum_{n=0}^{3} (-1)^{n} \tilde{v}^{n}}{y^{2d+6}} \, S_{d-2}(-3)\frac{\sqrt{\pi}(3d+17)\Gamma(\frac{d-1}{2})}{\Gamma(\frac{d+6}{2})} \, .
\end{align}

The upshot of the computations performed in this appendix is that, except when
$\ell=1$ for $I_{JJ\mathcal{O}}$, for generic dimensions and spins the surface
integrals does not vanish. Combined with the results using the current
conservation-condition shown in Appendix \ref{sec:appendix-A}, this enables to
show that at the higher-spin point the OPE coefficients of these three-point
function are trivial in Section \ref{sec:HS-at-infinite-distance} .

\subsection*{Integrals of $\langle J_\ell K_{3}\mathcal{O}\rangle$ Structures}

In Section \ref{sec:weakly-broken}, the charge-conservation identites leading
to $C_{JK\mathcal{O}}^\text{HS}=0$ can be simplified if some of the correlator
can be shown to be trivial. We focus now on correlators of the type $\langle
J_\ell K_{3}\mathcal{O}\rangle$ with $\ell\leq3$. In that case the conformal
structures at the positions given in equation \eqref{integral-point-choice}
are:
\begin{equation}
		\widehat{\Theta}_n = \frac{\widehat{P}_n}{|x_{12}|^{d-2}|x_{13}|^{d-2}|x_{23}|^{d+2}} = \frac{(-1)^{n+3-\ell}(\mathbf{x}^2-\frac{1}{4}y^2)^3}{y^{d+5-\ell}(\mathbf{x}^2+\frac{1}{4}y^2)^{d+1+\ell}}\,.
\end{equation}
The numerator of the conformal structures is given in equation \eqref{numerator-P-hat}. In the integrated Ward identity, with the choice of polarization
\eqref{polarization-choice}, the relevant integral over the surface $\Sigma$
given by $\mathbf{x}=0$ is therefore:
\begin{equation}
		I_{JK_3\mathcal{O}}(\ell) =\sum_{n=0}^{\ell}\int_\Sigma d^{d-1}\mathbf{x}\, u^n\widehat{\Theta}_n = 
		\frac{\sum_{n=0}^{\ell} (-1)^{n+3-\ell} u^n }{y^{d+5-\ell}} \int d^{d-1}\mathbf x 
		\frac{(\mathbf{x}^2-\frac{1}{4}y^2)^3}{(\mathbf{x}^2+\frac{1}{4}y^2)^{d+1+\ell}}\,,
\end{equation}
where the coefficients $u_n$ are those given in equation
\eqref{conservation-result-jk3O}. Once again, using spherical coordinates, it
can be written in terms of hypergeometric functions that further simplify to:
\begin{equation}
		I_{JK_3\mathcal{O}}(\ell) = 2^{d-4+2\ell}
		\frac{\sum_{n=0}^{\ell} (-1)^{n+2-\ell} u^n }{y^{2d+2+\ell}} S_{d-2}(\ell-1) (\ell^2+\ell+3(d-1))\frac{\Gamma(\frac{d-1}{2})\Gamma(\frac{d-3}{2}+\ell)}{\Gamma(d+1+\ell)}\,.
\end{equation}
This integral therefore never vanishes except when $\ell=1$, and we can then
follow the discussion around equation \eqref{JJO vanishes} to apply the
integrated Ward identity. Since the operator $K_3$ cannot appear in the
commutator $[Q_\ell,\mathcal{O}]$ when $\ell\leq3$ according to equation
\eqref{[Q,O]-[Q,K]}, the correlator must therefore be trivial when $\ell=2,3$,
which leads to a simplification of the charge-conservation identity
\eqref{charge-conservation-final} for spin-four currents.

\subsection*{Integrals of $\langle J_2 K_{\ell}\mathcal{O}\rangle$ Structures}

We focus now on correlators of the type $\langle
J_2 K_{\ell}\mathcal{O}\rangle$. In that case the numerator of  the conformal
structures is given in equation \eqref{numerator-p-prime}, and at the positions given in equation \eqref{integral-point-choice}, we obtain:
\begin{equation}
		\Theta^{\prime}_{n} = \frac{P^{\prime}_{n}}{|x_{12}|^{d-2}|x_{13}|^{d-2}|x_{23}|^{d+2}} = \frac{(-1)^{n+\ell}(\mathbf{x}^2-\frac{1}{4}y^2)^{\ell}}{y^{d+\ell}(\mathbf{x}^2+\frac{1}{4}y^2)^{d+\ell}}\,.
\end{equation}
In the integrated Ward identity, with the choice of polarization
\eqref{polarization-choice}, the relevant integral over the surface $\Sigma$
given by $\mathbf{x}=0$ is therefore:
\begin{equation}
		I_{J_2 K\mathcal{O}}(\ell) =\sum_{n=0}^{\min(\ell,2)}\int_\Sigma d^{d-1}\mathbf{x}\, u^n \Theta^{\prime}_{n} = 
		\frac{\sum_{n=0}^{\min(\ell,2)} (-1)^{n+\ell} u^n }{y^{d+\ell}} \int d^{d-1}\mathbf x 
		\frac{(\mathbf{x}^2-\frac{1}{4}y^2)^{\ell}}{(\mathbf{x}^2+\frac{1}{4}y^2)^{d+\ell}}\,,
\end{equation}
where the coefficients $u_n$ are those given in equation
\eqref{conservation-result-j2kO}. Once again, using spherical coordinates, it
can be written in terms of hypergeometric functions that further simplify to: 
\begin{equation}
		I_{J_2 K \mathcal{O}}(\ell) =
		\frac{\sum_{n=0}^{\min(\ell,2)} (-1)^{n+\ell} u^n }{y^{d+\ell}} \int d^{d-1}\mathbf x 
		\frac{(\mathbf{x}^2-\frac{1}{4}y^2)^{\ell}}{(\mathbf{x}^2+\frac{1}{4}y^2)^{d+\ell}}\,,
\end{equation}
where the integral can once again be written in terms of a hypergeometric function:
\begin{equation}
    \int d^{d-1}\mathbf x 
		\frac{(\mathbf{x}^2-\frac{1}{4}y^2)^{\ell}}{(\mathbf{x}^2+\frac{1}{4}y^2)^{d+\ell}} =(-1)^{\ell} \frac{2^{d}}{y^{d+1}}S_{d-2}\frac{\Gamma(\frac{d-1}{2})\Gamma(\frac{d-1}{2} + \ell)}{\Gamma(d-3 + 2 \ell)}\,{}_2F_1\left(\frac{d-1}{2}, -\ell, -\frac{d -7+4\ell}{2}; -1\right)
\end{equation}
We therefore find that the integral never vanishes for $\ell = 1,\dots,6$, and we can then
follow the discussion around equation \eqref{JJO vanishes} to apply the
integrated Ward identity. Since the operator $K_s$ cannot appear in the
commutator $[Q_2,\mathcal{O}]$ when $\ell \geq 1$ and when the spin two current is the energy-momentum tensor, $J_2 = T_2$, the correlator must therefore be trival when $\ell=1,\dots,6$,
which leads to a simplification of the charge-conservation identity
\eqref{charge-conservation-final} for spin-four currents.

\newpage
\bibliography{references}{}
\bibliographystyle{utphys}

%
\end{document}